\documentclass[pra,reprint,aps,superscriptaddress,floatfix,nofootinbib]{revtex4-2}

\usepackage{amsmath,amssymb,amsfonts}
\usepackage{graphicx}
\usepackage{bm}
\usepackage{xcolor}
\usepackage{placeins}
\usepackage{hyperref}

\begin{document}

\title{Intensity-noise suppression due to longitudinal mode coupling in
multimode intracavity-doubled lasers}

\author{Thomas~M.~Baer}
\affiliation{Department of Applied Physics, Stanford University,
Stanford, California 94305}

\date{\today}

\begin{abstract}
Multimode intracavity frequency-doubled lasers can reach states of
intensity-noise suppression orders of magnitude beyond the predictions
of independent-mode partition statistics. I show that the
$\chi^{(2)}$ coupled-wave dynamics in the doubling crystal obey a
dissipation inequality: a fourth-moment functional of the intracavity
intensity decreases monotonically under each forward nonlinear pass,
driving the mode ensemble toward a constant-intensity manifold. This
per-pass result is a strict Lyapunov inequality, and round-trip
simulations show attraction to the same low-$M_4$, nearly
constant-intensity state under the full cavity dynamics. I confirm the mechanism in an intracavity
frequency-doubled \mbox{Nd:YVO$_4$/LBO} laser, observing a
$100\times$ contrast between full and Fabry-Perot-filtered output
noise at fixed detector bandwidth, well beyond the $\sqrt{N/N_f}$
statistical-averaging baseline.  The argument depends only on the
algebraic structure of the coupling: a coherent superposition of
oscillators that share a quadratic dissipative channel.  It therefore
points to a general Lyapunov route to intensity-noise suppression in
coupled oscillator systems, beyond the intracavity-doubled laser.
\end{abstract}

\maketitle


\section{Introduction}

Intracavity frequency doubling is a standard way to generate
continuous-wave visible light from an infrared solid-state laser: a
nonlinear crystal placed inside the laser cavity converts the
circulating infrared field to its second harmonic with high
efficiency.  When such a laser oscillates on a small number of
longitudinal modes, it often exhibits large-amplitude intensity
fluctuations, the ``green problem''.  Early theoretical treatments of
intracavity doubling used single-longitudinal-mode rate-equation
models; the lasers they described operated in other regimes, but the
models did not treat the coupling among the multiple simultaneously
oscillating longitudinal
modes~\cite{smith1970,kennedybarry1974,barrykennedy1977,volosov1975,rice1978}
and did not explain the green problem.  A rate-equation treatment
extended the formalism to the multimode case by explicitly including
sum-frequency generation among the longitudinal modes, which was
identified as the primary cause of the intensity
instability~\cite{baer1986}.  The resulting few-mode dynamics, namely
bistability, mode-hopping, and deterministic chaos, were subsequently
explored by several
groups~\cite{wumandel1987,wiesenfeld1990,roybracikowskijames1991,erneuxmandel1995,kozyreffmandel1998,pietrzykdanailov2000}
and were early examples of optical bistability and optical chaos.

In this paper I model the opposite regime of the same system: when the
laser oscillates on hundreds of longitudinal modes, the total output
intensity becomes remarkably quiet~\cite{nighan1995}.  A striking
experimental signature is a large contrast between the full output and
the fluctuations of its individual modes.  A Fabry-P\'erot-filtered
few-mode subset is noisy, of order 45\% RMS, while the full output at
the same detector bandwidth is quiet, of order 0.4\% RMS.  This is a
contrast of approximately $100\times$, far beyond the $\sqrt{N}$
reduction expected from incoherent averaging of $N$ independent modes.
Here too, the earlier intensity-only rate-equation models did not
account for the quiet operation: the cancellation depends on the
relative phases of the modes, which intensity-only equations discard,
so a phase-resolved treatment is required.

I show, using a phase-resolved model of the coupled mode amplitudes,
that this quieting arises from the same longitudinal-mode coupling that
produces the fluctuations and instability of the few-mode case.  The
doubling crystal couples the modes through sum-frequency generation, as
in the few-mode case, but in the high-mode-count regime this coupling
organizes the amplitudes and phases of the modes so that their
individual fluctuations, which it in fact increases, cancel in the
total intensity.  The individual modes are loud; their sum is quiet.

The model I develop tracks the amplitude and phase of hundreds of
longitudinal modes.  It is an Ikeda-map iteration~\cite{ikeda1979} with
a two-step Fourier treatment of the coupled-wave $\chi^{(2)}$
interaction, and it includes spatial hole burning and back-conversion
explicitly.  The intensity-stabilization mechanism has a compact
analytic interpretation.  A fourth-moment measure of the intracavity
intensity, $M_4 = \langle I^2\rangle/\langle I\rangle^2$, decreases
monotonically under each forward pass through the crystal: brighter
time-slices are depleted more strongly than dim ones, through both
increased nonlinear loss and the phasing of the longitudinal modes,
flattening the intensity toward its mean.  This is a scalar Lyapunov
inequality for the forward pass; it drives the modes toward the
constant-intensity state that minimizes $M_4$.  I prove it analytically
for the fundamental field on a single forward pass with no incident
second harmonic.  The full round-trip dynamics, including saturable
gain, spatial hole burning, back-conversion, and noise, approach the
same stable total-intensity state numerically, where $M_4 \approx 1$.

Passive intensity-noise reduction has recently been studied in
single-frequency second-harmonic generation, where the nonlinearity
acts as an intensity-dependent loss that clamps the fluctuations of a
single field~\cite{shg_noise_porat2024,shg_power_zheng2024}; related
single-frequency schemes use nonlinear or quantum resources to filter
phase noise or generate squeezed
light~\cite{laser_phasenoise_li2026,squeezed_khz_li2025}.  The
mechanism here is different in kind, a classical multimode effect: the
nonlinearity \emph{increases} each mode's fluctuations and quiets the
output only through their collective cancellation, an effect with no
apparent single-mode analogue.  A fuller account of this history is
given in Appendix~\ref{app:prior}, covering early single-mode
intracavity doubling, the rate-equation treatments, the relation to the
FM laser, and the contrast with single-frequency nonlinear quieting.

In Sec.~\ref{sec:experiment} I describe the experiment: the
intracavity-doubled laser, its total-intensity stability, the two
detection paths, and the apparatus for measuring the amplitude
fluctuations of a filtered subset of longitudinal modes.
Section~\ref{sec:model} presents the numerical model.
Section~\ref{sec:lyap} gives a proof that $M_4$ decreases monotonically
for a single forward pass through the crystal, and shows numerically
that $M_4$ also descends under the full round-trip model as the laser
reaches a constant-intensity state.  Section~\ref{sec:results} compares
the predicted mode-count dependence of the instabilities with the
measurements and examines the geometric organization of the quiet
state by principal-component analysis.  Section~\ref{sec:discussion}
discusses the noise-cancellation mechanism and its generalization to
other coupled-oscillator systems that bear similarities to the laser
studied here.

\section{Experiment}\label{sec:experiment}

\begin{figure*}[tb]
\centering
\includegraphics[width=0.85\textwidth]{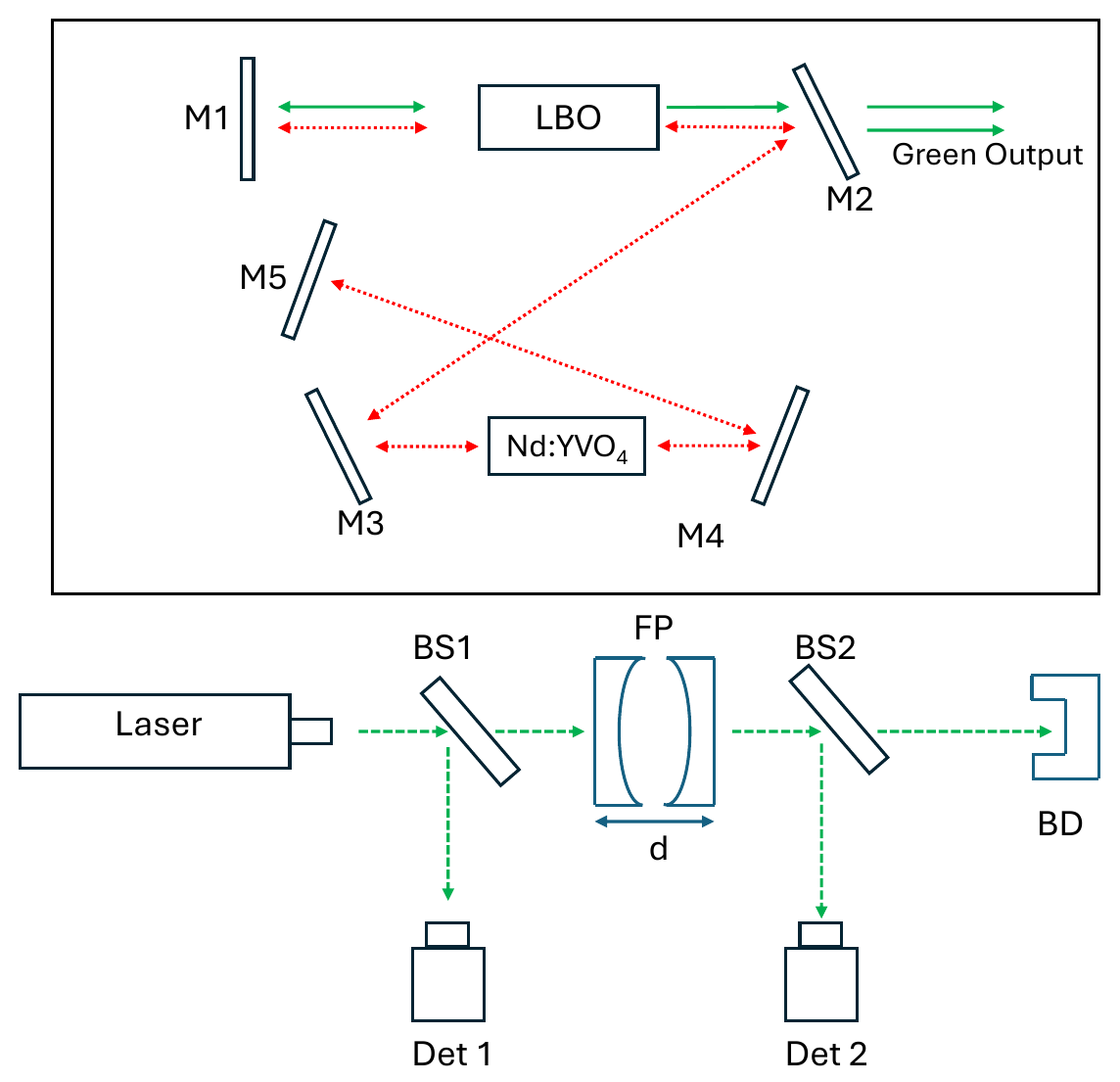}
\caption{Experimental layout.  \emph{Top:} the intracavity-doubled
laser, a five-mirror folded standing-wave cavity.  The Nd:YVO$_4$ gain
medium occupies the central arm (between folds M3 and M4); the LBO
doubling crystal sits in the arm next to end mirror M1, so the
circulating infrared (dotted red) double-passes the LBO each round
trip.  M1 and M5 are the cavity end high reflectors (HR at 1064~nm);
M1 is additionally HR at 532~nm, so the green (solid) generated in the
LBO is retroreflected back through the crystal and combines with the
forward-generated green.  M2 is a dichroic fold mirror (HR at 1064~nm,
transmitting at 532~nm) that couples the green output out of the
cavity; M3 and M4 are folds (HR at 1064~nm).  \emph{Bottom:} detection.
The green output is split by BS1 into a full-output detector (Det~1)
and a path through a free-standing Fabry-Perot etalon (FP, mirror
spacing $d$, finesse $\approx 75$) that transmits an $N_f$-mode subset
to a second detector (Det~2) via BS2; the residual beam is dumped (BD).
Det~1 (full output) and Det~2 (Fabry-Perot-filtered output) are
identical: same photodiode, transimpedance amplifier, and layout, with
measured bandwidth $\ge 1$~MHz, so the full-versus-filtered contrast is
instrument-independent.  Varying $d$ selects $N_f \approx 3, 6, 15, 34$
transmitted modes.}
\label{fig:setup}
\end{figure*}

\subsection{The experimental laser platform}

The experimental laser is a commercial intracavity-doubled
diode-pumped solid-state laser, the Spectra-Physics Millennia~Vs, a
representative example of the now-standard commercial architecture for
multimode intracavity-doubled lasers.
The gain medium is Nd:YVO$_4$ in a five-mirror folded standing-wave
cavity (Fig.~\ref{fig:setup}, top), with the gain crystal in the
central arm and the LBO doubler in the arm next to an end mirror, so
the infrared double-passes the doubler each round trip and the green is
retroreflected and coupled out through a dichroic fold mirror.  The
doping density and crystal-cut orientation are typical for this class
of laser.  The doubling crystal is intracavity LBO, phase-matched for
type~I noncritical phase matching at the operating wavelength.  The
pump is a multimode fiber-coupled diode array.  Total green output
reaches 5~W at 532~nm.  The cavity supports approximately 200
longitudinal modes across the gain bandwidth, spaced by the 358~MHz
intermode beat.  Data was collected at the typical operating point of
approximately 4~W green output.

\subsection{Detection: full output and Fabry-Perot-filtered output}

Two detection paths are used (Fig.~\ref{fig:setup}, bottom).  The first
samples the full green output at 532~nm with a fast photodiode and a 1~MHz
analog detector bandwidth.  This band covers the frequency region where
the laser's technical intensity noise predominates and where it is most
detrimental to applications that use the laser as a pump source: in
optical-frequency-comb generation, for example, amplitude fluctuations
of the frequency-doubled pump in this band are a known source of
carrier-envelope and frequency noise in the pumped
comb~\cite{matos2006,sutyrin2013}.  It also
lies far below the 358~MHz intermode beat, so the measurement reports the
baseband envelope-intensity noise of the total field rather than the
mode-beat structure.  The
second passes the green output through a free-standing
Fabry-Perot etalon used as a tunable spectral selector.  The etalon
mirrors have 4~cm radius of curvature and are mounted on adjustable
rails; the mirror spacing is set independently of the laser cavity,
allowing the etalon free spectral range $\mathrm{FSR}_{\rm FP}$ to be
varied while the finesse $\mathcal{F} \approx 75$ remains fixed by
the mirror reflectivities.  The transmission peak width is then
$\Delta\nu_{\rm FP} = \mathrm{FSR}_{\rm FP}/\mathcal{F}$, and the
number of Millennia modes within each peak is
$N_f \approx \Delta\nu_{\rm FP}/358~\mathrm{MHz}$.  I use four
configurations spanning $N_f \approx 3, 6, 15, 34$ modes, selected
by adjusting the mirror spacing.  The etalon lies entirely outside the
laser cavity: it neither selects nor controls which longitudinal modes
oscillate, but passively samples a contiguous $N_f$-mode window of the
already-formed green field.  The oscillating mode count ($N \approx 200$)
is fixed by the gain bandwidth and cavity length, not by any adjustable
element; the laser free-runs at that count while I vary only the
downstream etalon passband.  Both detection paths feed identical
detectors with identical 1~MHz analog bandwidths.  The contrast
between full and filtered outputs at the same detector bandwidth is
the primary observable.  The reported RMS relative intensity noise is
the standard deviation of a detector signal normalized to its mean,
$\sigma_P/\langle P\rangle$, evaluated over each 5~ms record after
low-pass filtering to the 1~MHz analog bandwidth; the traces are
digitized at 10~MS/s and no detrending is applied.

\subsection{Primary measurement: full output vs FP-filtered}\label{sec:vbprimary}

\begin{figure*}[tb]
\centering
\includegraphics[width=0.80\textwidth]{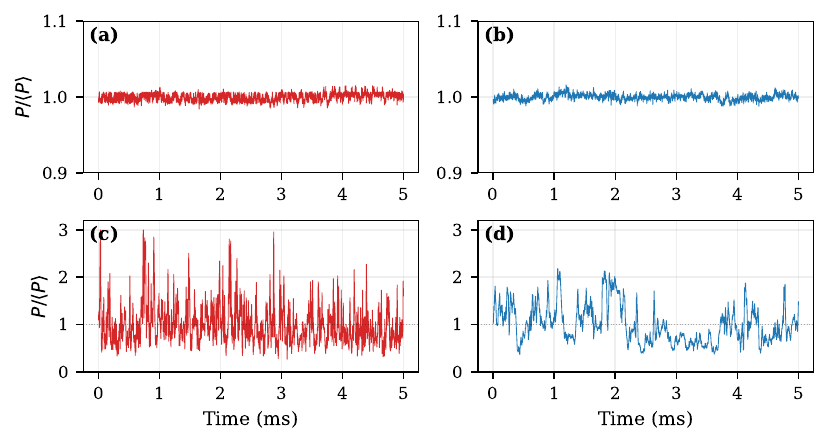}\\[6pt]
\includegraphics[width=0.62\textwidth]{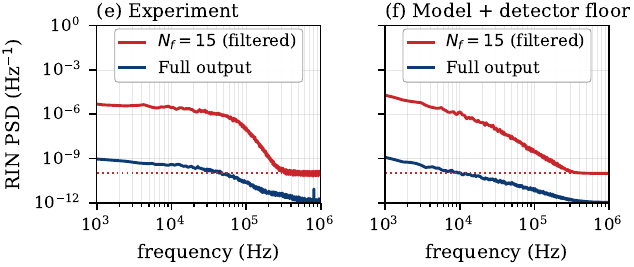}
\caption{Full green output versus a Fabry-Perot-filtered few-mode
subset, experiment and model side by side.  \textbf{(a)--(d)~Time
domain} (model parameters in Table~\ref{tab:params}): the full Millennia
green output, (a)~experiment and (b)~simulation; and the same field
after an FP etalon (finesse $\approx 75$) selecting an $N_f \approx 15$
mode subset, (c)~experiment and (d)~simulation.  Full-output RMS
$0.44 \pm 0.02\%$ (exp) / $0.42 \pm 0.03\%$ (sim); filtered-output RMS
$\approx 45\%$ (exp) / $\approx 38\%$ (sim); the filtered RMS is a
chaotic single-realization quantity that varies by of order $30\%$
run-to-run, so these are representative values.  The full-output error
bars are standard deviations over thirteen experimental 5~ms windows and
44 simulation seeds; the measured $\sim 100\times$ contrast is far beyond any
independent-averaging baseline, which predicts a reduction of only a
factor of a few ($\sqrt{N_{\rm eff}/N_f} \sim 2$).
\textbf{(e),(f)~Relative intensity noise:} RIN power
spectral density of the full output (navy) and the $N_f = 15$ filtered
subset (red), (e)~experiment and (f)~model plus measurement floor;
dotted lines are the measurement noise floor.  The two detectors are
identical (same photodiode, transimpedance amplifier, and layout;
measured bandwidth $\ge 1$~MHz), so the full-versus-filtered contrast is
instrument-independent, and the additive floor differs between
channels only through the lower transmitted intensity of the filtered
channel.}
\label{fig:fp}
\end{figure*}

Figure~\ref{fig:fp} shows the result at the typical operating point.
The full green output runs at $0.44 \pm 0.02\%$ RMS intensity noise
within the 1~MHz analog detector bandwidth, and a Fabry-Perot etalon
($N_f \approx 15$, finesse $\approx 75$) selecting a spectral subset of
the same field runs at about $45\%$ RMS.  The filtered output is a
chaotic few-mode signal whose RMS varies by of order $30\%$ from capture
to capture, so this value is representative rather than a precise
reproducible number; the full-output RMS above, by contrast, is a robust
attractor property (mean and standard deviation over the thirteen
nonoverlapping 5~ms windows of the 70~ms record).  The simulation, using a single $1/f$ gain-noise
channel calibrated to the full-output RMS ($\sigma_{\rm gain}
= 5\times10^{-6}$, spectral slope $-1.08$, update interval $\Delta n
= 50$ round trips $\approx 140$~ns), reproduces both panels:
$0.42 \pm 0.03\%$ full and about $38\%$ filtered across the 44 noise
seeds, the filtered value varying by of order $30\%$ from seed to seed,
with power spectral density agreement across the frequency
decades below the technical-noise shoulder and no parameter
retuning between full and filtered outputs.  The $1/f$ gain perturbation
is the only adjustable technical-noise process in the model;
spontaneous emission remains as a fixed quantum-floor source
(Appendix~\ref{app:noise}), and the technical process is represented by
fifty partially common-mode spatial streams.  The model includes no
separate parameters for pump, thermal, or mechanical fluctuations.

The contrast between full and filtered output is approximately
$100\times$ at this representative passband ($N_f \approx 15$).  The
$\sqrt{N/N_f}$ statistical-averaging baseline depends on the
appropriate definition of the active mode count.  Four counts enter the
comparisons that follow: the full oscillating band ($N \approx 200$),
the active count ($N_{\rm active} \approx 86$, modes whose mean power
exceeds $1\%$ of the peak), the power-weighted effective count
($N_{\rm eff} \approx 65$, defined below), and the number of
Fabry-P\'erot-transmitted modes ($N_f$, from $3$ to $34$, with
$N_f \approx 15$ representative).  The mean intracavity
power is concentrated in modes whose individual time-averaged powers
vary substantially: weighting modes by their mean intensity gives the
effective mode number $N_{\rm eff} = (\sum_i \langle P_i\rangle)^2 /
\sum_i \langle P_i\rangle^2 \approx 65$ at the typical operating
point.  $N_{\rm eff}$ is the equivalent number of equally bright
independent modes that would produce the same total power as the
actual unequal mode-power distribution; for an ensemble of
identically bright modes $N_{\rm eff}$ equals the mode count, and
for a distribution dominated by a single mode $N_{\rm eff}$
approaches one.  This is the appropriate count for an
incoherent-averaging
estimate, smaller than either $N_{\rm active} \approx 86$
(Sec.~\ref{sec:psisweep}, modes whose mean power exceeds 1\% of the
peak) or the full $N \approx 200$ band.  Even this conservative
independent-mode count gives a statistical-averaging reduction of only
order $\sqrt{N_{\rm eff}/N_f} \sim 2$, far short of the measured
$\sim 100\times$ contrast.  Incoherent averaging is therefore not the
mechanism.  The result is consistent with the coherent cancellation
predicted by the Lyapunov analysis: the same noise drives both the
full and the filtered output, and the full output is quiet because
the modes' contributions to the total intensity cancel.
Figure~\ref{fig:nfsweep} (Sec.~\ref{sec:vc-modecount}) presents the
cancellation as a function of $N_f$ across four passband settings,
demonstrating that the contrast varies smoothly with the filter width
and is not specific to one operating point.

Several alternative explanations for the full-versus-filtered contrast
are bounded by the experimental geometry.  The full-output and
filtered-output detectors operate at the same 1~MHz analog bandwidth
and share the same source field, removing the bandwidth-dependent
and source-dependent artifacts.  The Fabry-Perot is parked at a fixed
position during each measurement, with cavity-length stability
sufficient that the passband does not drift through the modes during
the recording.  Frequency-to-amplitude conversion through passband
slope is bounded by adjusting the mirror spacing to maximize
transmission (parking the etalon on a transmission peak rather than
on a slope) and verifying that small intentional detuning produces
consistent RMS values.  Pointing-induced amplitude
modulation is bounded by aperturing the beam upstream of the etalon.
Detector linearity is verified by comparing fixed-attenuation traces.
The simulation uses a single calibrated $1/f$ gain-noise channel with
no phase-noise injection.  It reproduces the contrast quantitatively
across the full sweep of Fig.~\ref{fig:nfsweep}.  This is further
evidence that the measurement is dominated by the multimode coherent
cancellation rather than by filter-side artifacts.

\section{Model}\label{sec:model}

\subsection{Physical setup and coupled-wave dynamics}\label{sec:coupledwave}

The system is a standing-wave solid-state laser with a homogeneously
broadened gain medium~\cite{sargent1974,siegman1986} and an intracavity
nonlinear crystal
phase-matched for second-harmonic generation.  The cavity supports
hundreds of longitudinal modes spaced by the free spectral range
$\Omega = \pi c / L_{\rm cav}$, all competing for gain through spatial
hole burning in the standing-wave gain grating and coupled through the
nonlinear crystal.  I focus on the typical operating point: pump rates
well above threshold, single-pass conversion efficiencies in the
few-percent range, and output coupling dominated by the nonlinear
crystal rather than a partially transmitting mirror.  I treat the
linear standing-wave configuration, the most common commercial design;
the $\chi^{(2)}$ mode coupling that drives the stabilization is not
specific to it, but the spatial-hole-burning grating and the
double-pass geometry are.

The intracavity infrared field is written as a sum over longitudinal
modes,
\begin{equation}
E(t,z) = \sum_k E_k(z)\, e^{i\omega_k t},
\label{eq:field-expansion}
\end{equation}
with $\omega_k = \omega_0 + k\Omega$ and complex mode coefficients
$E_k(z)$ tracked explicitly in amplitude and phase.  The mode index
$k$ runs over a range that extends well beyond the gain bandwidth
(typically $|k| \le 250$, giving 501 modes total).  The number of
modes that actually oscillate is determined self-consistently by the
gain dynamics rather than fixed a priori.  The second-harmonic field
$E_{\rm grn}(t,z)$ is expanded similarly in Fourier components
$E_{{\rm grn},m}(z)$ at frequencies $2\omega_0 + m\Omega$.  Tracking
the phase of each mode, not only its intensity, is essential for
describing the coherent back-conversion channel of the $\chi^{(2)}$
interaction, which couples mode phases across the ensemble.

Within the crystal, the mode coefficients evolve under the
coupled-wave equations
\begin{align}
\frac{dE_k}{dz}            &= i\kappa \sum_j E_{{\rm grn},k+j}\, E_j^*,\nonumber\\
\frac{dE_{{\rm grn},m}}{dz} &= i\kappa \sum_{j+k=m} E_j\, E_k,
\label{eq:cwe}
\end{align}
where $\kappa$ is the nonlinear coupling coefficient.  These equations
describe forward conversion and back-conversion on equal footing.
Intensity-rate-equation treatments approximate this interaction as a
lumped loss proportional to $|E|^2$, which captures the
amplitude-equalization channel but discards the coherent phase
coupling through which mode phases are organized.

\subsection{Numerical implementation}

The model is implemented as an Ikeda-map iteration.  Each round trip
is the composition of three operators acting on the full mode vector
$\{E_k\}$:
\begin{equation}
E_k^{(n+1)} = \hat{N}\bigl(\hat{C}\bigl(\hat{G}(E_k^{(n)})\bigr)\bigr) + \xi_k^{(n)}.
\label{eq:roundtrip}
\end{equation}
Here $\hat{G}$ is the gain stage, $\hat{C}$ is the $\chi^{(2)}$
crystal stage, and $\hat{N}$ is the linear cavity-loss stage.  The
term $\xi_k^{(n)}$ is stochastic drive injected once per round trip
per mode, modeling spontaneous emission and technical noise.

The gain stage evolves a dynamical, frequency-dependent saturated gain
$g_i$ and applies it to the field amplitude together with the
round-trip loss:
\begin{equation}
\tau_f\,\frac{dg_i}{dt} = g_0(\nu_i) - s_i\,g_i, \qquad
E_i \to E_i\,\exp\!\left[(g_i - \ell)/2\right],
\end{equation}
where $g_0(\nu_i)$ is the Lorentzian small-signal gain, $\ell$ the
round-trip linear loss, $\tau_f$ the gain-recovery time, and
$s_i = 1 + \sum_j \beta_{ij}|E_j|^2/I_s$ the saturation factor with
$I_s$ the saturation power.  In the adiabatic limit $\tau_f \to 0$
the gain relaxes instantly to $g_i = g_0(\nu_i)/s_i$, recovering the
quasi-static saturated form; at the physical $\tau_f = 90~\mu$s it lags
the field, which is the origin of the slow mode-partition drift
identified below.  The full equations, including the Lorentzian gain
profile that sets the operating mode bandwidth, are in
Appendix~\ref{app:model}.  The coefficient
$\beta_{ij} = \beta(|i-j|)$ is the standing-wave spatial-hole-burning
cross-saturation between modes $i$ and $j$.  For a gain medium of
fractional length $\eta = L_{\rm rod}/L$ centered at axial position
$z_0$,
\begin{equation}
\beta(m) = \frac{2}{3}\left[1 + \frac{1}{2}\,
              {\rm sinc}(m\pi\eta) \cos(2 m \pi z_0/L)\right],
\end{equation}
with $m = |i-j|$.  The sinc envelope limits SHB competition to modes
within $\Delta m \sim 1/\eta$ of each other.  The cosine factor
selects which mode separations couple strongly.  Mode separations
beyond the sinc envelope decouple.  This geometric selection sets the
active mode count $N_{\rm active}$ as a function of $z_0$.

The crystal stage $\hat{C}$ is computed in the time domain using a
two-step Fourier method.  The time-domain field $E(t)$ is constructed
by inverse FFT from the input mode vector.  The coupled-wave equations
of Sec.~\ref{sec:coupledwave} are then integrated through the forward pass.  The
high-reflector boundary condition $\delta\psi$ is applied, and the
back-pass equations are integrated.  Here $\delta\psi$ is the relative
phase shift between the IR and the back-traveling green field acquired
on reflection from the high-reflector at the end of the crystal arm.
It is set by the high-reflector coating and the air path.  In
commercial intracavity-doubled lasers it is engineered to maximize
green output coupling.  The output
IR field is projected back onto mode amplitudes by forward FFT.  This
split-step approach captures all pairwise four-wave-mixing
interactions among modes exactly within the mode basis.  The cost is
$O(N \log N)$ per round trip, enabling simulation of $N \sim 200$
modes over multimillisecond timescales.

Two approximations in the model deserve explicit comment.
Intracavity dispersion is neglected.  The free spectral range is
taken uniform across the gain bandwidth, which is justified for
short-cavity solid-state lasers where dispersive mode-frequency
shifts are small compared to the intermode beat.  Upper-state
spatial diffusion, which in principle washes out the standing-wave
gain grating, is also neglected.  For the solid-state gain media
modeled here, diffusion over the upper-state lifetime is well below
the grating period, and SHB is effectively static on the relevant
timescales.  Neither approximation measurably affects the dynamics
in the regime of interest.

The full parameter set used in all simulations, the noise model, and
convergence tests are given in Appendix~\ref{app:model}.

\subsection{The fourth-moment ratio $M_4$}

The quantity central to the analysis is the fourth-moment ratio of
the total intracavity intensity,
\begin{equation}
M_4 \equiv \frac{\langle I(t)^2 \rangle}{\langle I(t)\rangle^2},
\quad I(t) = \Bigl|\sum_k E_k\, e^{i\omega_k t}\Bigr|^2,
\label{eq:m4-def}
\end{equation}
where the time average is over the round-trip period.  $M_4 \ge 1$
with equality if and only if $I(t)$ is constant.  A fully uncorrelated
multimode field with Gaussian-speckle statistics gives $M_4 = 2$.
$M_4$ is the Lyapunov candidate for the theorem below.

The value $M_4 = 1$ defines the ideal constant-intensity limit.  A
finite set of modes approaches but does not attain this limit exactly:
an exactly constant-modulus multimode field requires an unbounded
sideband series (Appendix~\ref{app:fm}), so a finite mode band settles
at a small residual $M_4 - 1 > 0$.  Throughout, I refer to the resulting
low-$M_4$ states, and to the constant-intensity set they approach, as
the near-constant-intensity manifold, and use ``the $M_4 = 1$ manifold''
as shorthand for it.

The theorem-level observable is the IR intracavity intensity $I(t)$.
The experimental observable in Figs.~\ref{fig:fp} and \ref{fig:nfsweep}
is the green output $P_{\rm grn}$.  In the weak-conversion limit the
time-averaged generated green power is proportional to
$\langle I(t)^2\rangle = M_4\,\langle I\rangle^2$, which makes the link
between the forward-pass Lyapunov functional and the measured output
transparent: at fixed circulating power a decrease in $M_4$ is a
proportional decrease in green power, so the green RMS tracks $M_4$
excursions monotonically (the two are not numerically equal, because of
the $\langle I\rangle^2$ normalization in $M_4$).  The simulations do
not rely on this approximation.  They propagate the complex green field
through the full forward- and backward-pass coupled-wave equations, and
the simulated detector signals are computed directly from the green
spectrum (Appendix~\ref{app:model}), with the measured Fabry-Perot
transfer function applied for the filtered channel.  I use green RMS as
the operational figure of merit when comparing to experiment, and
IR-side $M_4$ for theorem verification.

\section{The Lyapunov structure}\label{sec:lyap}

\subsection{The descent statement}

As the field propagates through the crystal, $M_4$ decreases
monotonically with propagation distance $z$, with equality only when
$I(t)$ is constant:
\begin{equation}
\frac{dM_4}{dz} \le 0, \qquad z \in [0, L_{\rm crystal}].
\label{eq:descent-body}
\end{equation}
The rigorous proof for the forward pass through the crystal is given
in Appendix~\ref{app:proof} and holds throughout the operating regime
of laboratory intracavity-doubled lasers, including peak single-pass
conversion efficiencies well in excess of typical commercial designs.
The extension to the back pass under the standard commercial
double-pass design is treated in Appendix~\ref{app:proof} as an
assumption, supported numerically by the $\delta\psi$-sweep simulation
of Sec.~\ref{sec:psisweep} and by the experimental measurements
presented below.

The intensity-channel mechanism is intuitive.  The crystal depletes
intensity as $I^2$, so brighter time-slices lose proportionally more
energy than dim time-slices, flattening $I(t)$ toward its mean.  The
phase channel operates by a different mechanism, through coherent
four-wave-mixing torques on the mode phases; this channel is described
in Sec.~\ref{sec:two-channels}.

Structurally, the result is analogous to the monotone approach to
thermodynamic equilibrium: a scalar quantity evolves monotonically
under the dynamics until it reaches an extremum.  Boltzmann's
H-theorem in kinetic theory is the canonical example.  Here $M_4$
plays the role of $H$, decreasing under the crystal dynamics toward
the constant-intensity state that minimizes it.  The analogy is formal,
not thermodynamic: it is a monotone-descent structure, and implies no
temperature, entropy, or ensemble.

The laser is modeled as a system of several hundred coupled nonlinear
mode equations.  $M_4$ condenses their collective evolution into the
descent of a single scalar toward the low-noise, constant-intensity
state.  One scalar suffices because $M_4$ is a function of the total
intensity.  That intensity responds to fluctuations in both the
individual mode intensities and their relative phases.  These are the
amplitude and phase channels described in Sec.~\ref{sec:two-channels}.

The same scalar tracks how the system responds to its physical
parameters.  The number of operating modes is set by the cavity length
and the gain bandwidth.  When it is reduced, the descent can no longer
overcome the gain dynamics.  $M_4$ then rises above unity, and the
system leaves the manifold for the bistable and chaotic regimes of
few-mode operation (Appendix~\ref{app:prior}).  The placement of the
gain medium acts through the spatial-hole-burning grating.  It selects
among the on-manifold states and admits the quasi-FM configuration
(Appendix~\ref{app:fm}).  In this sense $M_4$ is more than a proof that
the quiet regime exists.  It is an order parameter for its stability.

The consequence is that $M_4 = 1$, the constant-intensity state, is
the per-pass descent target of the crystal dynamics, regardless of
initial mode amplitudes, phases, or count.  Whether the round-trip composition
reaches this manifold depends on the balance between the in-crystal
descent and the inter-pass dynamics (gain saturation, SHB, and noise
injection), which is treated in Sec.~\ref{sec:roundtrip} and verified
numerically in Sec.~\ref{sec:results}.

\subsection{Why the descent holds}

The descent has a transparent physical origin.  In the weak-conversion
limit, the $\chi^{(2)}$ interaction depletes each time slice of the
field in proportion to its own intensity squared,
$\delta I(t) \propto -\varepsilon\, I(t)^2$, where $\varepsilon$ is the
dimensionless coupling strength.  Bright slices lose more than dim ones.
Every crystal pass therefore flattens the intensity waveform and lowers
$M_4$.

A short calculation makes this exact at leading order.  To first order
in $\varepsilon$, $\Delta M_4 \le 0$ follows from the $-I^2$ depletion
and a Cauchy-Schwarz inequality on the round-trip time average, with
equality only when $I(t)$ is constant.  The result assumes nothing
about mode count, amplitudes, or phases.  The calculation is given in
Appendix~\ref{app:leading}.

This leading-order argument establishes the descent directly from the
time-domain intensity-dependent depletion, without separating its
diagonal and off-diagonal contributions in the mode basis.  The
amplitude-equalization and phase-organization channels of
Sec.~\ref{sec:two-channels} are a physical interpretation of that
nonlinear map.  A full proof, valid to all orders, is
given in Appendix~\ref{app:proof}.  It uses the exact coupled-wave
solution and supplies a peak-conversion bound.  The bound places
laboratory intracavity-doubled lasers safely inside the regime where
the descent holds.

\subsection{Two channels: amplitude equalization and phase organization}
\label{sec:two-channels}

The $\chi^{(2)}$ interaction in the mode basis decomposes into two
structurally distinct channels acting on every crystal pass.  When
$E(t)|E(t)|^2$ is expanded in modes, each mode receives contributions
from triplets of modes $(j, k, l)$.  A diagonal triplet is one in
which the same mode appears as both raised and lowered index.
Concretely, these are terms of the form $E_j |E_k|^2$ that contribute
to mode $j$.  The remaining contributions, in which three distinct modes
combine to drive a fourth, are the four-wave-mixing triplets.  The
amplitude-equalization channel arises from the diagonal triplets,
producing intensity-quadratic depletion proportional to each mode's
own amplitude.  This channel cooperates with spatial-hole-burning
cross-saturation in the gain medium and drives the mode-amplitude
envelope toward a smooth gain-shaped form.  It does not by itself
organize phases.

The phase-organization channel arises from the four-wave-mixing
triplets, which are off-diagonal in the mode basis and produce phase
shifts on individual modes proportional to the sine of four-mode phase
combinations.  Each mode $i$ receives a torque from every triplet
$(j, k, l)$ satisfying the frequency-matching condition
$j + k - l = i$, with the torque vanishing when the phase combination
$\varphi_i - \varphi_j - \varphi_k + \varphi_l$ is zero.  The
mode-basis decomposition therefore suggests a phase-organizing channel
through these off-diagonal four-wave-mixing terms; in simulations,
this channel drives the field toward phase configurations with
strongly suppressed coherent intensity peaks, consistent with the
$M_4 = 1$ manifold of Sec.~\ref{sec:lyap}.

The structural form of this torque is opposite to that of Kuramoto
coupling \cite{strogatz2000}, which uses an attractive sinusoidal
force to align oscillators toward a mean-field phase.  The
four-wave-mixing torque is repulsive.  It drives mode $i$ away from configurations in which its
phase aligns with others, because alignment maximizes the intensity
peaks that the $\chi^{(2)}$ depletion preferentially attacks.  A
natural class of low-$M_4$ configurations under this torque resembles
a splay-phase arrangement in which the modes are phased so that
constructive interference at any single instant is suppressed
\cite{silber1993}.  This is a structural analogy, not a formal
reduction: where the attractive Kuramoto coupling drives synchrony, the
repulsive $\chi^{(2)}$ torque drives the modes to null the coherent sum
rather than to align with the mean field.

\subsection{Round-trip dynamics and approach to the manifold}
\label{sec:roundtrip}

Between crystal passes, the gain medium amplifies each mode according
to its gain and the spatial-hole-burning coupling, which in general
does not respect the Lyapunov descent.  The observed steady-state
$M_4$ is the round-trip balance between the descent in the crystal
and the drive from the gain medium and noise sources.  The magnitude
of the steady-state residual $M_4 - 1$ is set by this balance and is
determined numerically.

\begin{figure*}[tb]
\centering
\includegraphics[width=\textwidth]{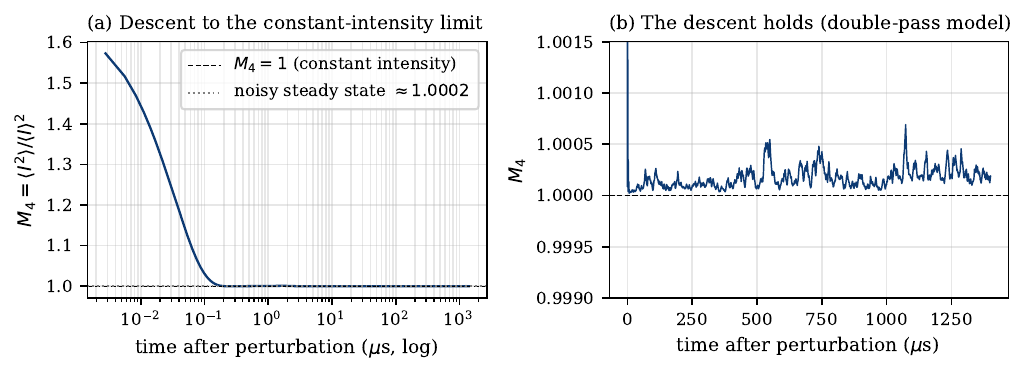}
\caption{Numerical descent of $M_4 = \langle I^2\rangle/\langle
I\rangle^2$ under the full double-pass round-trip model (canonical
42~cm configuration, $\sigma_{\rm gain} = 5\times10^{-6}$).  At $t=0$ a
concentration perturbation raises $M_4$ to $\approx 1.7$.
\textbf{(a)}~On a logarithmic time axis, $M_4$ descends to the
constant-intensity limit $M_4 = 1$ within $\sim 0.2~\mu$s.
\textbf{(b)}~Over the full $1.4$~ms record the descent holds: $M_4$
fluctuates at the noisy steady-state residual $\approx 2\times10^{-4}$
above one, with no drift back toward the perturbed value.  The analytic
proof (Appendix~\ref{app:proof}) covers the single forward crystal
pass; this figure is the numerical evidence that the descent survives
the full round-trip composition.}
\label{fig:m4transient}
\end{figure*}

At the operating point, a perturbation that concentrates the
intracavity intensity into the central few modes raises $M_4$ to
$\approx 1.7$.  Under the full
double-pass round trip (Fig.~\ref{fig:m4transient}) $M_4$ descends to the
constant-intensity limit
within a fraction of a microsecond and then holds at the noisy
steady-state residual $\langle M_4\rangle - 1 \approx 2\times10^{-4}$
for the entire $1.4$~ms record, with no drift back toward the perturbed
value.  The descent proved for one forward crystal pass therefore
persists, and is stable, under composition with the gain step, the
back-reflected pass, and noise.

The same descent holds across a broad band of the round-trip phase
$\delta\psi$, failing only in the runaway band near $\delta\psi = \pi$
(Sec.~\ref{sec:psisweep}).

Two further structural features of the descent are essential for what
follows.  First, the theorem establishes that the per-pass change in
$M_4$ from the crystal is nonpositive in the operating regime.
Whether the system actually reaches the manifold depends on whether
the inter-pass gain dynamics are dynamically subordinate to the
crystal's descent.  When subordinate, the system relaxes toward $M_4
= 1$.  When the gain medium pumps $M_4$ back up enough to balance the
crystal at nonstationary $M_4$, the round-trip composition produces
a stable limit cycle off the manifold.  The regimes of Fig.~\ref{fig:regimes}
(Appendix~\ref{app:prior}) and the quiet state of Fig.~\ref{fig:quiet} are organized by
which side of this balance dominates.  Second, $M_4$ is invariant under uniform scaling of the intensity, so
it is insensitive to the overall power level.  This is the counterpart
of the loss nonlinearity: the flattening is driven by the super-linear,
$\propto I^2$ crystal loss, which depletes bright time-slices more than
dim ones, whereas a purely linear loss would deplete every slice in
equal proportion and merely rescale $I(t)$.  A slow drift in the overall
level---e.g.\ the level change from adiabatic gain dynamics on the
fluorescence timescale $\tau_f \gg \tau_c$---therefore leaves $M_4$
unchanged; the descent organizes the relative mode amplitudes and phases
that set the shape of $I(t)$, not the overall level.

The theorem applies to the full-bandwidth intensity $I(t)$ within the
round-trip period.  Experimental measurement with a detector
bandwidth slow compared to the intermode beat frequency samples the
slow-envelope $M_4$, which is related to but distinct from the
full-bandwidth quantity.  Both are meaningful indicators of the mechanism.  The slow-envelope
$M_4$ inferred from the measured RMS, $M_4 = 1 + (\sigma_P/\langle
P\rangle)^2$, is far below the Gaussian-speckle limit $M_4 = 2$; the
full-bandwidth $M_4$, which the detector does not resolve, lies well
below $2$ in the simulations.

\subsection{The quiet multimode regime}\label{sec:attractors}

\begin{figure}[tb]
\centering
\includegraphics[width=\columnwidth]{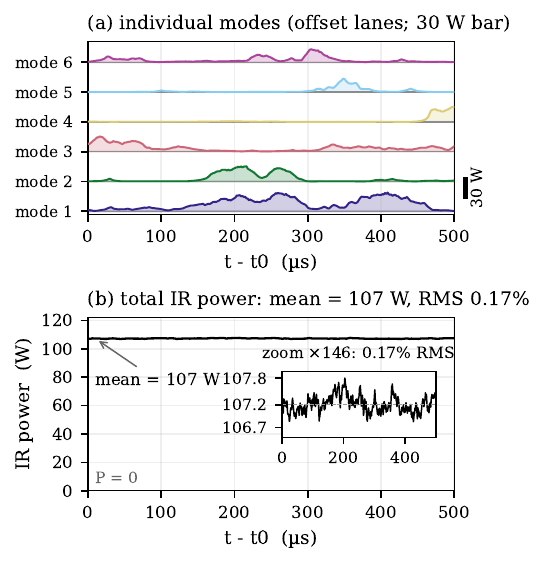}
\caption{Quiet multimode regime ($N \approx 200$, 42~cm Millennia
geometry, $z_0 = 0.025$~m, $\sigma_{\rm gain} = 5\times10^{-6}$,
seed~42).  \textbf{(a)}~The six most active longitudinal modes, drawn
in offset lanes so the individual traces stay legible; each mode is
shaded down to its own lane baseline (that mode's zero), and every mode
swings over $\sim 0$--$34$~W.  The vertical scale is common to all lanes
and is set by the 30~W bar at right.  \textbf{(b)}~The total IR power
$P_{\rm IR} = \sum_j |E_j|^2$ on a true absolute scale: $P = 0$ sits at
the bottom of the frame and the mean, $\langle P_{\rm IR}\rangle
\approx 107$~W, is a nearly flat line near the top.  The inset zooms
the same trace by $\sim\!150\times$ to expose the residual, which is
held to only 0.17\% RMS.  The individual modes are loud while their sum
is quiet; this is the intensity-noise cancellation.  The doubled green output RMS is
0.39\% for this seed (44-seed mean $0.42 \pm 0.03\%$,
Sec.~\ref{sec:vbprimary}).}
\label{fig:quiet}
\end{figure}

Figure~\ref{fig:quiet} shows the intracavity dynamics of the
high-mode-count quiet state ($N \approx 200$) in the 42~cm Millennia
geometry.  The individual longitudinal modes [panel~(a), each drawn in
its own offset lane] fluctuate strongly: the per-mode circulating power
$|E_j(t)|^2$ swings over $\sim 0$--$34$~W on the timescale shown.  The
total intracavity IR power [panel~(b), on a true absolute scale from
$P = 0$] holds $\langle P_{\rm IR}\rangle \approx 107$~W with an RMS of
only 0.17\%, resolved in the inset zoom.  The simulated green output for this seed runs at 0.39\% RMS,
near the 44-seed mean of $0.42 \pm 0.03\%$ (Sec.~\ref{sec:vbprimary}),
larger than the IR by roughly the factor of two of the
intensity doubling.  This is the regime that motivates the paper.

The same model also reproduces the historical few-mode regimes of
intracavity-doubled lasers, studied over four decades.  It uses the
same Lyapunov structure, the same standing-wave spatial-hole-burning
kernel, and the same coupled-wave crystal physics.  These regimes are
bistable mode-hopping and chaotic sequential
pulsing at $N = 2$--$3$, and quasi-FM operation.  These are shown in
Appendix~\ref{app:prior} (Fig.~\ref{fig:regimes}).  What selects the
regime is the balance between the gain dynamics and the in-crystal
descent: when the descent dominates the system relaxes onto the
constant-intensity manifold (the quiet and quasi-FM states); when the
gain dynamics overpower it, the system settles into an off-manifold
limit cycle (the bistable and chaotic states).

\section{Comparison and geometric organization}\label{sec:results}

\subsection{Mode-count dependence}\label{sec:vc-modecount}

A second test varies the Fabry-Perot passband width across four mode
counts.  The same simulation parameters as Fig.~\ref{fig:fp} are
used; only the FP passband changes.  Each mode-count column is a
separate experimental capture at a different etalon spacing, taken
with a different detector acquisition than Fig.~\ref{fig:fp}.  The
absolute filtered RMS therefore differs between the two figures at
the nominal $N_f \approx 15$ setting.  What is comparable is the
trend in filtered noise with $N_f$, not the RMS at any single
passband.

\begin{figure*}[tb]
\centering
\includegraphics[width=\textwidth]{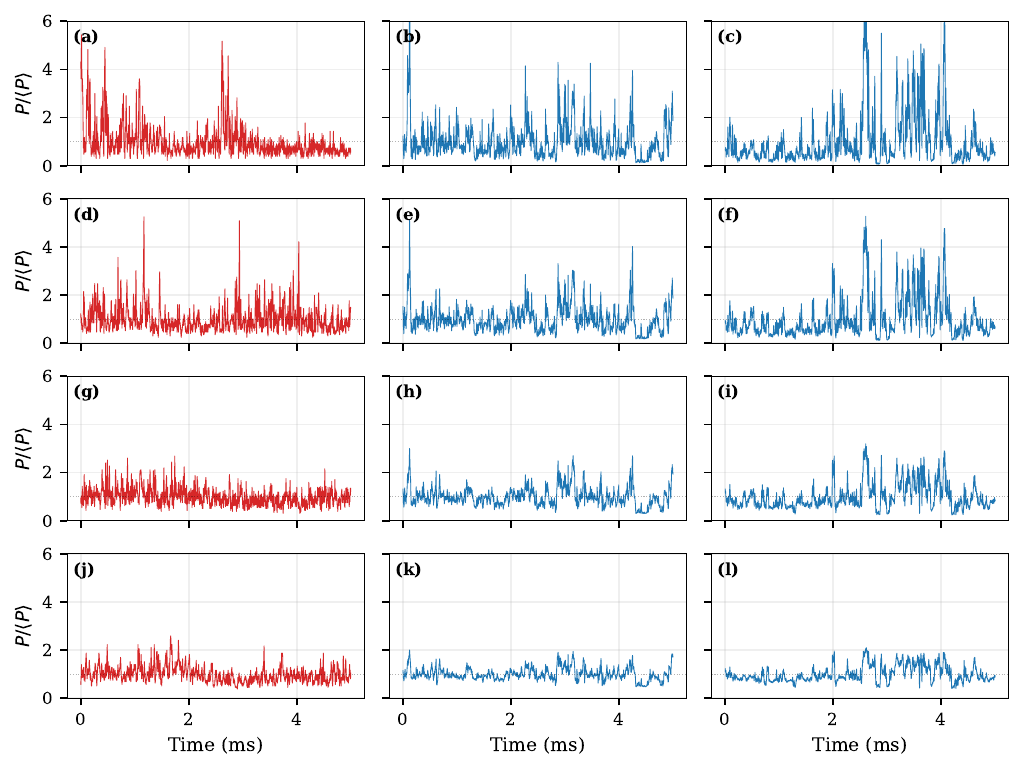}
\caption{Time-domain traces of FP-filtered output power, normalized
to mean power.  Panels~(a, d, g, j) show experimental measurements
at Fabry-Perot passband settings corresponding to $N_f \approx 3,
6, 15$, and $34$ transmitted longitudinal modes, respectively.
Panels~(b, e, h, k) and (c, f, i, l) show simulation realizations
at the same passband settings from two different noise seeds,
illustrating trajectory-to-trajectory variation comparable to what
different oscilloscope trigger conditions capture experimentally.
At narrow passbands (top rows), both experimental and simulation
traces exhibit the characteristic spike-and-plateau intermittency
of few-mode dynamics; at wider passbands (bottom rows), this
intermittency averages out into smoother variability approaching
the full-output noise floor.  The qualitative phenomenology is
reproduced across simulation realizations; quantitative measures of
agreement (multiseed statistics) are given in Sec.~\ref{sec:vc-modecount}.}
\label{fig:nfsweep}
\end{figure*}

Figure~\ref{fig:nfsweep} shows the progression.  At $N_f \approx 3$
the filtered output is dominated by spike events reaching
$\sim 5\times$ the mean, with long quiet intervals between them;
the experimental trace runs at 68.9\% RMS over the displayed 5~ms
window, and the two simulation seeds (137, 9001) run at 73\% and
107\%, bracketing the experiment.  At $N_f \approx 6$ the spikes
become more frequent and moderate (exp 53.5\%, sim 56-84\%).  At
$N_f \approx 15$ the noise transitions to a more distributed
character (exp 34.1\%, sim 40-53\%).  At $N_f \approx 34$ the
output is nearly flat and sim and experiment agree closely (exp
33.0\%, sim 26-34\%).  The seed-to-seed scatter in the simulation
is a chaotic-trajectory effect: individual realizations differ at
the 10-30\% level even when ensemble statistics are tightly
constrained.

The progression connects the few-mode and high-mode regimes within a
single mechanism.  The few-mode panels expose the differential mode
fluctuations that the full output cancels.  As $N_f$ grows, the
contributions to the filtered intensity that the $\chi^{(2)}$ descent
has organized to cancel in the full output begin to cancel in the
filtered output as well.  The $N_f \approx 34$ panels already show
partial cancellation, and the trend extrapolates smoothly to the
0.42\% RMS at $N \gtrsim 200$ shown in Fig.~\ref{fig:fp}.

\subsection{Green output and noise suppression co-optimize with the round-trip phase}
\label{sec:psisweep}

The high-reflector phase $\delta\psi$ between the IR and the
back-traveling green field fixes the relative phase at which the
back-pass $\chi^{(2)}$ interaction resumes.  It is set by the
high-reflector coating and the air path between the crystal and the
mirror.  A commercial
laser is tuned to maximize green output, which corresponds to
constructive interference of the forward-generated and back-reflected
green at the entrance of the back pass; in the convention used here
that optimum is $\delta\psi = 0$.

The forward-pass dissipation theorem of Appendix~\ref{app:proof} is
independent of $\delta\psi$: it concerns a single crystal pass with no
incident green.  The phase enters only through the back pass, where the
phase-rotated forward green re-enters the crystal and drives
back-conversion into the IR.  To see how the intensity stabilization
depends on this phase I swept $\delta\psi$ from $0$ to $2\pi$ in the
full cascaded round-trip model.  This model carries the green field
through the reflection rather than merely combining it at the output.
I held all other parameters fixed and averaged over four noise
realizations.

\begin{figure*}[tb]
\centering
\includegraphics[width=\textwidth]{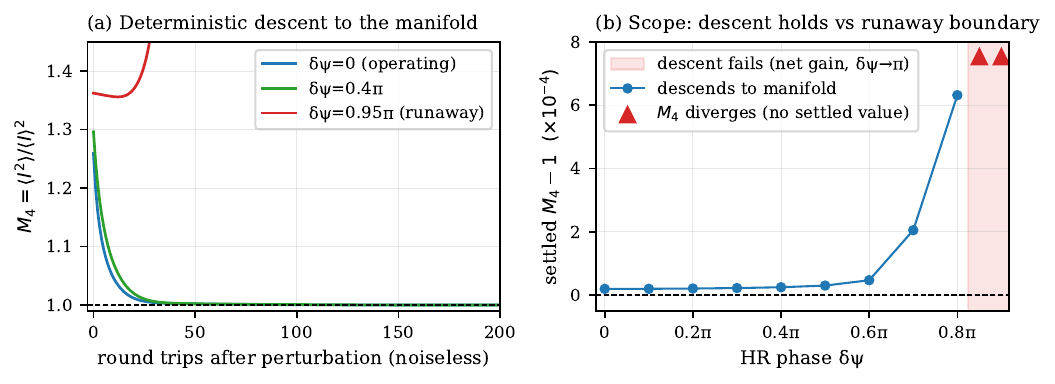}
\caption{Dependence of the $M_4$ descent on the round-trip phase
$\delta\psi$, under the full round-trip model with gain, spatial hole
burning, and $\chi^{(2)}$ back-conversion.  \textbf{(a)}~Deterministic
(noiseless) descent of $M_4(t)$ after a phase perturbation, for three
high-reflector phases: the operating point $\delta\psi = 0$, a
representative $\delta\psi = 0.4\pi$, and $\delta\psi = 0.95\pi$ in the
runaway band.  The descent is monotone at the operating and
representative phases; near $\delta\psi = \pi$ the back-pass
$\chi^{(2)}$ is net parametric gain and $M_4$ diverges.
\textbf{(b)}~Settled $M_4$ versus $\delta\psi$: the descent reaches the
manifold across a broad band around the operating point and fails only
near the runaway boundary. Triangles in the shaded band mark phases
where $M_4$ diverges and no settled value exists.}
\label{fig:m4descent}
\end{figure*}

The green output and the IR intensity stabilization are optimized by
the same phase.  Both are best at $\delta\psi = 0$, where the green
output is maximal and the residual $\langle M_4\rangle - 1 =
3.6\times10^{-5}$ (total IR RMS $0.05\%$ in this cascaded model, of the
same order as the operating-point residual reported above).  As
$\delta\psi$ increases the two degrade together:
across the stabilized band the green output and
$\log(\langle M_4\rangle - 1)$ track with correlation $-0.99$.  A phase
excursion that costs a few percent of green output is accompanied by a
larger but still small rise in the residual.  The residual reaches
$\sim\!7\times10^{-4}$ (IR RMS $0.17\%$) by $\delta\psi \approx
0.83\pi$.  Near $\delta\psi = \pi$ the back-reflected green re-enters
in antiphase: the back-conversion becomes net parametric gain and the
intracavity field runs away over a narrow band
($0.85\pi \lesssim \delta\psi \lesssim 0.96\pi$), and at
$\delta\psi = \pi$ exactly the two green contributions cancel, the
crystal extracts no green, and the multimode intensity noise returns
($\langle M_4\rangle \to 2$).  The map is symmetric about $\pi$.
Figure~\ref{fig:m4descent} shows the deterministic descent at three
representative phases and the settled $M_4$ across the sweep.

This co-optimization is not a coincidence.  The same return-pass
nonlinear interaction that transfers power into the green field is the
dissipation that drives the IR toward the constant-intensity state, so
strong constructive coupling maximizes both at once.  There is no trade-off
between green extraction and intensity stabilization: a laser tuned for
maximum green output sits, by construction, at the phase of tightest
cancellation, and with wide margin, since the green output is nearly flat
over the range where the residual stays far below its uncancelled value.

At the operating phase the time-averaged intensity is carried by
roughly 86 longitudinal modes, defined as those whose mean intensity
over the record window exceeds 1\% of the time-averaged peak.  This
subset accounts for 99.1\% of the mean IR power, the remaining modes
within the gain bandwidth carrying residual leakage.  The roughly 200
modes referenced in Fig.~\ref{fig:quiet} and elsewhere are the full set
the simulation evolves above quantization noise, of which the 86 are
the dynamically participating subset.

\subsection{Geometric organization revealed by principal-component analysis}
\label{sec:pca}

The Lyapunov descent identifies $M_4 = 1$ as the constant-intensity
manifold targeted by the per-pass crystal dynamics; in the simulated
round-trip dynamics the system approaches this limit, so the total
intracavity intensity $I(t)$ approaches a constant.  The descent
statement does not say how the system organizes its mode-amplitude
fluctuations to achieve that.  The geometric structure of the steady
state can be exposed by
a principal-component analysis of the per-mode intensity time series,
and that structure is a feature of how the descent operates that goes
beyond the bare $M_4 \to 1$ statement.

Writing
$\delta I_i(t) = I_i(t) - \langle I_i\rangle$ for the fluctuation of
mode $i$ about its mean and $\bm{\delta I} = (\delta I_1,\dots,\delta I_N)$
for the fluctuation vector, the covariance matrix of $\{\delta I_i\}$
diagonalizes into principal components $\mathbf{v}_n$ with time-domain
projections $\xi_n(t) = \mathbf{v}_n \cdot \bm{\delta I}(t)$, each with an
intrinsic fluctuation amplitude (the eigenvalue $\lambda_n$, the variance
of $\xi_n$) and a squared overlap $c_n^2 = (\mathbf{v}_n\cdot\mathbf{u})^2$
with the uniform direction $\mathbf{u} = (1,\dots,1)/\sqrt{N}$, the
direction of the total intracavity power.  The
variance of the total power is
$\mathrm{var}(P_{\rm tot}) = N\sum_n c_n^2\,\lambda_n$
(Appendix~\ref{app:pca}), so a principal component contributes to the
total-output noise only if it both fluctuates and points along
$\mathbf{u}$.

The steady state separates into two subspaces.  The principal
components that build $\mathbf{u}$ are individually quiet, their
projections fluctuating at the milliwatt level; the loud principal
components, which carry watt-level fluctuations, are nearly orthogonal
to $\mathbf{u}$.  The total-power variance is only $\sim 6\times10^{-5}$ of the value it
would have if the modes fluctuated independently---a suppression of four
orders of magnitude.  The origin is geometric: the loud fluctuations are
held out of the total by their orthogonality to $\mathbf{u}$, a coherent
cancellation, not the $\sqrt{N}$ reduction of independent averaging.

\begin{figure}[tb]
\centering
\includegraphics[width=\columnwidth]{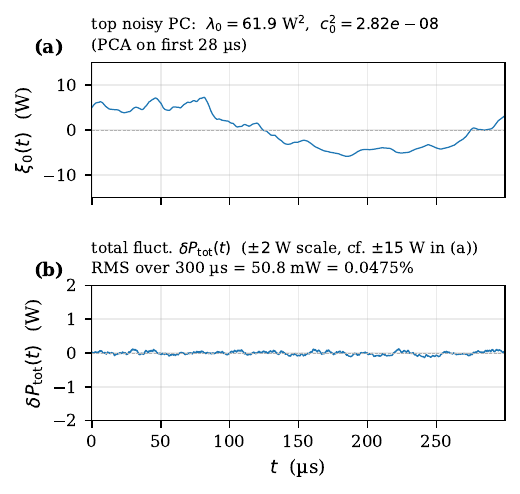}
\caption{The mode makeup of the total intensity drifts while the total
stays quiet.  Principal-component analysis of the simulated per-mode
intensity fluctuations (seed~42): the PCA basis is computed on a
$10^4$-pass ($\approx 28~\mu$s) window at the converged steady state,
and both panels show a $300~\mu$s record projected onto the dominant
noisy eigenvector $\mathbf{v}_0$ from that window ($\lambda_0 =
61.9$~W$^2$, overlap $c_0^2 = 2.82\times10^{-8}$ with $\mathbf{u}$).
\textbf{(a)}~The projection $\xi_0(t) = \mathbf{v}_0\cdot\bm{\delta I}(t)$
wanders over several watts across the record: the loud subspace is
nonstationary, and the fixed short-window eigenvector slowly loses its
grip as the noisy direction drifts on $\sim\tau_f$.  \textbf{(b)}~The
total intracavity-power fluctuation $\delta P_{\rm tot}(t)$, on an
expanded $\pm 2$~W scale (cf. $\pm 15$~W in (a)), shows only a small
residual (RMS $50.8$~mW, $0.0475\%$ of $\langle P\rangle$, more than two
orders of magnitude below the $\approx 7.9$~W RMS of the dominant
principal component).  The contrast between the drifting loud
projection and the nearly pinned total is the signature of the
mechanism: the system diffuses across the constant-intensity manifold
while the total observable stays quiet.}
\label{fig:pca}
\end{figure}

The decomposition also shows what the constant-intensity manifold means
dynamically, and this is the main reason for the analysis.  The loud
subspace is not stationary.  The specific mode combinations that carry
the noise drift and reorganize continuously on the gain-recovery time
$\tau_f$: the leading noisy direction computed over one window is nearly
orthogonal to the one computed a millisecond later.  The system diffuses
slowly across the high-dimensional flat $M_4 = 1$ manifold while the
total output stays clamped at every instant.  The invariant of the quiet state is therefore not any particular set of
modes---the loud and quiet subspaces both drift---but the quietness of
the total output itself, a collective cancellation that can be
characterized as dynamic cloaking.  Figure~\ref{fig:pca}
shows this directly.  Across the $300~\mu$s record the dominant noisy
direction $\xi_0(t)$ wanders over several watts as the loud subspace
drifts (a), while the total fluctuation $\delta P_{\rm tot}(t)$ shows
only a small residual, more than two orders of magnitude smaller in
RMS (b).  The same quiet
total-power direction, and the same $\sim\tau_f$ drift of the loud
subspace, appear for every noise seed and analysis window I examined;
the covariance decomposition and the window-length dependence of the
loud subspace are given in Appendix~\ref{app:pca}.

The damping of the total-intensity fluctuations is strongest for
components slow compared to the e-folding time ($\sim 280$~ns at typical operating parameters,
corresponding to roughly 100 round trips), where this e-folding time
is the timescale over which $M_4 - 1$ decreases under the cumulative
round-trip composition.  The residual fluctuation in $P_{\rm tot}$ is
dominated by faster components that the cumulative descent has not had
time to organize across the available passes.  The mechanism therefore
has a bandwidth: slow large fluctuations are effectively suppressed in
the total intensity, while higher-frequency residual components survive.

This organization is consistent with the phase-organization
channel of Sec.~\ref{sec:two-channels}.  The four-wave-mixing torques
drive the modes toward splay-like phase arrangements in which
collective fluctuations along $\mathbf{u}$ are suppressed.
Fluctuations in directions orthogonal to $\mathbf{u}$ are not
constrained by the descent.  They remain free to carry the bulk of the
fluctuation energy.

\section{Discussion}\label{sec:discussion}

The defining feature of this mechanism sets it apart from earlier routes to
quiet nonlinear output.  It does not quiet the individual longitudinal modes;
it makes them noisier.  Activating the crystal raises the per-mode intracavity
relative intensity noise by roughly an order of magnitude, from a few percent
to tens of percent.  At the same time the noise of the total intracavity
intensity, and with it the green output, falls.  The enlarged per-mode
fluctuations are cancelled in the aggregate: the modes organize into an
anticorrelated, total-conserving configuration whose sum stays quiet.  Without
the crystal the modes fluctuate independently, and the total is suppressed only
by statistical ($\sqrt{N}$) averaging.  With it the cancellation is coherent,
and it exceeds $\sqrt{N}$ by more than an order of magnitude.  The experimental
signature is the large full-versus-filtered contrast of the green output
(Fig.~\ref{fig:fp}).  The full output collects the whole mode ensemble, so the
anticorrelated per-mode fluctuations cancel and it is quiet.  The
Fabry-P\'erot filter passes only a few of the modes; it therefore samples their
large individual fluctuations without the anticorrelated partners that cancel
them, and the filtered green is noisy.

Prior nonlinear routes to amplitude-stable or quiet doubled output work in
different ways.  The FM laser~\cite{harristarg1964}
imposes a constant-intensity Bessel state by external phase modulation.
Single-frequency intracavity doubling clamps the fluctuations of one field
through the intensity-dependent nonlinear
loss~\cite{shg_noise_porat2024,shg_power_zheng2024}, and related schemes filter
phase noise or generate squeezed
light~\cite{laser_phasenoise_li2026,squeezed_khz_li2025}.  None has a multimode
analogue of the collective cancellation described here.  The detailed relation
to the FM laser is given in Appendix~\ref{app:fm}, and to single-frequency
nonlinear quieting in Appendix~\ref{app:singlemode}.

The mode configuration that carries out this cancellation is not fixed.  As the
principal-component analysis shows (Sec.~\ref{sec:pca}), the weighted mode
combinations that momentarily carry the output-intensity fluctuations drift and
reorganize continuously.  The laser's mode makeup wanders over the
constant-total-intensity manifold on the gain-recovery timescale $\tau_f$, while
the total output stays clamped.  What is preserved is the
quietness of the total, not any particular set of modes.  This is the dynamic
cloaking of Sec.~\ref{sec:pca}.  The mechanism is correspondingly
basin-flexible: many distinct configurations satisfy the descent constraint with
$\mathbf{u}$ in the quiet subspace.  The one the laser occupies is a property of
its history, not a fine-tuned hardware feature.

This dynamic, collective character makes the quiet state resilient.  The
cancellation belongs to the whole ensemble, not to any fixed mode configuration.
A perturbation that disturbs individual modes---mechanical vibration, a thermal
or alignment drift---therefore does not defeat it; the system simply relaxes to
a different point on the same constant-intensity manifold.  The mechanism does
have a finite bandwidth.  It is set by the $\sim 280$~ns ($\sim 100$ round-trip)
e-folding time of the descent (Sec.~\ref{sec:pca}): fluctuations slow compared
to that time are cancelled from the total most effectively, faster residual
components less so.  Technical and mechanical noise is the dominant limitation of
real lasers, and the part most detrimental to pump applications.  It lies at
Hz-to-kHz frequencies, well within this band.  It is therefore precisely what
the mechanism suppresses, and it does so with no active stabilization.

The quiet state is reached in tension with spatial hole burning.  The
$\chi^{(2)}$ interaction supplies two coupling channels with opposite relations
to SHB.  Amplitude equalization, through the intensity-quadratic SHG loss,
cooperates with SHB cross-saturation and drives the amplitude envelope toward the
gain-shaped form.  Phase organization, carried by coherent back-conversion, can
instead fight SHB.  The phase relations that minimize $M_4$ are not in general
those that maximize SHB-driven gain extraction at the efficient operating point,
and the residual $M_4 - 1$ floor is set by this balance.  Which configuration is
selected is fixed by where the gain-SHB constraint surface intersects the
manifold: a quasi-FM state at the few-mode case of Fig.~\ref{fig:regimes}(c), a
gain-shaped envelope at the high-mode-count case of Fig.~\ref{fig:quiet}
(Appendix~\ref{app:fm}).  The crystal thus plays a dual role.  Numerical
simulations show that increasing $\varepsilon$ overcomes the SHB-driven mode competition that
would otherwise limit free-running Nd:YVO$_4$ operation to $\approx 26$ active
modes, allowing the much larger active populations seen with the crystal in
place.  At the same time it organizes the active-mode phases toward $M_4 = 1$.
Both effects activate well below commercial operating parameters.

Finally, the descent depends only on the algebraic structure of the coupling: a
coherent superposition of oscillators at distinct frequencies, and a dissipative
channel whose rate scales quadratically in their total amplitude.  It is
therefore not specific to intracavity-doubled lasers.  Globally coupled
oscillator arrays such as Josephson junctions~\cite{tsang1991} and coupled laser
arrays~\cite{fabiny1993,silber1993} share the multioscillator architecture and a
constant-total-amplitude manifold of equilibria, but they realize the loss in
different forms.  Whether a quadratic dissipative channel of the form required
here can be engineered into any of them is an open question.  Where it can, the
same route to intensity-noise suppression should follow.

\section{Conclusion}\label{sec:conclusion}

I have shown that a multimode intracavity frequency-doubled laser
reaches a state in which the total output intensity is far quieter than
the mode-partition noise of its individual longitudinal modes would
predict.  The mechanism is the monotone decrease of the fourth-moment
ratio $M_4$ under the coupled-wave $\chi^{(2)}$ dynamics in the
crystal, which drives the modes toward a constant-intensity state; the
individual modes remain loud, and their fluctuations cancel in the sum.

A phase-resolved numerical model that tracks the amplitude and phase of
every longitudinal mode reproduces the measurements quantitatively.  It
reproduces the full-versus-filtered contrast of the quiet regime, a
factor of approximately $100\times$, well beyond the $\sqrt{N/N_f}$
averaging baseline.  It captures the dependence of this contrast on the
number of transmitted modes.  It also reproduces the bistable, chaotic,
and quasi-FM regimes of few-mode operation, each with the cavity
parameters appropriate to it.  The analytic result
covers the single forward crystal pass with no incident second
harmonic; the full round-trip descent, including gain, spatial hole
burning, back-conversion, and noise, is numerical.

The same $\chi^{(2)}$ mode coupling that produced the green-problem
instabilities in few-mode operation nearly four decades
ago~\cite{baer1986} produces the intensity-noise suppression at high
mode count, once the complex mode amplitudes, including phase, are
retained.

\begin{acknowledgments}
The author thanks Professor Martin Fejer and Professor Stephen Harris
for useful discussions.  Anthropic's Claude and the Claude Code assistant
(Claude Opus~4.x family, 2026) were used to assist with editing and
organizing the text, with writing and refactoring the simulation and
figure-generation code, and with reference and consistency checking; the
author directed this use, verified all results and statements, and is
solely responsible for the content.  All figures are quantitative plots of
simulated or measured data and contain no AI-generated or AI-modified
imagery.
\end{acknowledgments}

\appendix

\section{Model details}\label{app:model}

The simulation treats each longitudinal mode as a complex field
amplitude $E_i(t)$ ($i = 1, \dots, N$), evolved per round trip
through three sequential physical stages: gain saturation including
spatial-hole-burning cross-saturation in the gain medium; nonlinear
coupling through the intracavity $\chi^{(2)}$ crystal in a
double-pass configuration; and output coupling.  Two noise mechanisms
are included: spontaneous emission as a quantum-limit floor injected
once per round trip, and technical noise modeling pump speckle,
thermal drift, and other engineering perturbations as $1/f$-filtered
fluctuations on the gain coefficient with partial-common-mode
mode-spatial structure (Sec.~\ref{app:noise}).  The cavity is a
standing-wave linear resonator, with Nd:YVO$_4$ as the gain medium and
LBO as the doubling crystal.  It uses the dichroic geometry of
commercial intracavity-doubled lasers: the high-reflector at the end of
the crystal arm reflects both the fundamental and the forward-generated
second harmonic back through the crystal, and the green output is
coupled out at a dichroic fold mirror (the output coupler).  All parameters used in the simulations are those
typical for the platform studied here, summarized at the end of this
section.

\subsection{Round-trip structure}

A single round trip $E_i^{(n)} \to E_i^{(n+1)}$ is the composition
\begin{equation}
E_i^{(n+1)} = \hat{N}\bigl[\hat{C}\bigl[\hat{G}[E_i^{(n)}]\bigr]\bigr]
              + \xi_i^{(n)},
\label{eq:composition}
\end{equation}
where $\hat{G}$ is the gain stage, $\hat{C}$ is the $\chi^{(2)}$
crystal stage, and $\hat{N}$ is the linear cavity loss including
output coupling.  Each operator acts on the full mode vector
$\{E_i\}$.  The additive term $\xi_i^{(n)}$ models spontaneous
emission injected once per round trip per mode.  Round-trip linear
cavity losses (residual output-coupler transmission, intracavity optic
absorption, scatter) are represented by the mode-independent loss
$\ell$ of the gain stage below.  In the typical operating-point design
that is the subject of the Lyapunov analysis, no frequency-selective
elements are present in the IR cavity beyond the doubling crystal
itself, so $\ell$ is mode-independent; as a common amplitude factor it
leaves $M_4$, which is scale-invariant, unchanged.
The historical-reproduction simulations of Fig.~\ref{fig:regimes}(a)--(c)
include frequency-selective elements as part of the experimental
designs being reproduced: the etalon used by Baer~\cite{baer1986} for
two-mode and three-mode selection in panels~(a) and~(b), and the
birefringent filter used by Tsunekane~\cite{tsunekane1997} for the
quasi-FM state in panel~(c).  These elements impose mode-dependent
losses specific to those configurations and are not the channel
responsible for the descent.  The dominant mode-dependent loss in
the typical-operating-point simulations is the nonlinear conversion
through the crystal.

\subsection{Gain stage with spatial-hole-burning}

The gain stage integrates a dynamical saturated gain and applies it,
together with the round-trip loss, to the field amplitude.  The peak
small-signal gain is frequency dependent, following the Lorentzian gain
line of the medium,
\begin{equation}
g_0(\nu_i) = \frac{g_0}{1 + \bigl[\,2(\nu_i - \nu_0)/\Delta\nu_g\,\bigr]^2},
\label{eq:A1a}
\end{equation}
with $\nu_0$ the line center, $\Delta\nu_g = 250$~GHz the gain FWHM, and
$g_0$ the peak round-trip small-signal gain.  The saturated gain $g_i$
of each mode relaxes toward its saturated value on the
fluorescence-recovery time $\tau_f$,
\begin{equation}
\tau_f\,\frac{dg_i}{dt} = g_0(\nu_i) - s_i\, g_i,
\qquad
s_i = 1 + \sum_j \frac{\beta_{ij}\,|E_j|^2}{I_s},
\label{eq:A1b}
\end{equation}
integrated once per round trip by an explicit Euler step of size
$\tau_c$, the round-trip time.  Here $s_i$ is the saturation factor,
$I_s$ the saturation power, and $\beta_{ij} = \beta(|i-j|)$ the
standing-wave spatial-hole-burning cross-saturation coefficient between
modes $i$ and $j$.  The gain then multiplies the field amplitude
together with the round-trip linear loss $\ell$,
\begin{equation}
E_i \to E_i\,\exp\!\bigl[(g_i - \ell)/2\bigr].
\label{eq:A1c}
\end{equation}
The loss $\ell$ enters only here, in the amplitude step; it is the
round-trip loss denoted $\hat N$ in the composition
(\ref{eq:composition}) and is not applied a second time.

In the adiabatic limit $\tau_f \to 0$ the gain reaches its saturated
fixed point instantaneously,
\begin{equation}
g_i = \frac{g_0(\nu_i)}{s_i},
\label{eq:A1}
\end{equation}
the quasi-static saturated gain.  At the physical $\tau_f = 90~\mu$s,
with $\tau_f \gg \tau_c$, the gain instead lags the field: each round
trip moves $g_i$ only a fraction $\tau_c/\tau_f \approx 3\times10^{-5}$
of the way to (\ref{eq:A1}).  This lag is the slow channel of the
dynamics.  It sets the timescale on which power is redistributed among
the modes and is the origin of the $\sim\tau_f$ drift of the mode
partition seen in the principal-component analysis
(Sec.~\ref{sec:pca}).  Modes far from line center,
$|\nu_i - \nu_0| \gtrsim \Delta\nu_g$, have small $g_0(\nu_i)$; when
their saturated gain falls below the loss $\ell$ they drop below
threshold and decay.  The gain bandwidth $\Delta\nu_g$ thus bounds the
number of modes above threshold, and together with the
spatial-hole-burning geometry below it fixes the active mode count
$N_{\rm active}$.  For a gain
medium of length $L_{\rm rod}$ centered at axial position $z_0$ in a
cavity of length $L$,
\begin{equation}
\beta(m) = \frac{2}{3}\bigl\{1 + \tfrac{1}{2}\,
                {\rm sinc}(m\pi\eta) \cos(2 m \pi z_0/L)\bigr\},
\label{eq:A2}
\end{equation}
with $\eta = L_{\rm rod}/L$ the gain-medium fractional length and
$m = |i-j|$.  The sinc envelope limits SHB competition to modes
within $\Delta m \sim 1/\eta$ of each other.  The cosine factor is
the position dependence that selects which mode separations couple
strongly.  Modes beyond the sinc envelope decouple from each other
and can coexist; this geometric selection sets the active mode count
$N_{\rm active}$ as a function of $z_0$ for given gain bandwidth.

\subsection{$\chi^{(2)}$ crystal stage in double-pass geometry}

The intracavity LBO crystal is operated in a double-pass
configuration: the IR field traverses the crystal in the forward
direction, the IR and forward-generated green are both reflected from
the high-reflector that closes the cavity arm, and the IR plus
reflected green re-enter the crystal traveling backward.  The
forward-pass and back-pass dynamics are governed by the coupled-wave
equations
\begin{equation}
\frac{dE}{dz}        = i\eta_c\, E_{\rm grn}\, E^*, \qquad
\frac{dE_{\rm grn}}{dz} = i\eta_c\, E^2,
\label{eq:A3}
\end{equation}
where $E(t)$ and $E_{\rm grn}(t)$ are the time-domain IR and green
fields and $\eta_c = \sqrt{\varepsilon}/L_c$ is the single-pass field
coupling coefficient with $\varepsilon \propto d^2_{\rm eff} L_c^2$ the
dimensionless coupling strength.  This $\eta_c$ is the single-field form
of the modal coupling $\kappa$ used in Sec.~\ref{sec:coupledwave} and
Appendix~\ref{app:exact}; the two are identical, $\kappa \equiv \eta_c$.  The forward-pass boundary condition
is $E_{\rm grn}(z=0,t) = 0$ at the crystal entrance.  The back-pass
boundary condition is $E_{\rm grn}(z=L_c,t) = E_{\rm grn}^F(L_c,t)
\exp(i\delta\psi)$, where $E_{\rm grn}^F$ is the green field at the
crystal exit at the end of the forward pass and $\delta\psi$ is the
relative phase shift between IR and green imposed by the
high-reflector coating and the air path between the crystal and the
mirror.

In the laser studied here, the high-reflector phase is engineered so
that $\delta\psi$ produces constructive interference of
self-second-harmonic generation on the back pass.  The back-reflected
green and the back-generated green add in phase.  They drive forward
conversion (IR $\to$ green) rather than back-conversion (green $\to$
IR).  This is the standard design choice for commercial
intracavity-doubled lasers, where $\delta\psi$ is tuned to maximize
the second-harmonic output.  Under this design, both passes deplete
IR and add to green; the back-pass green field acts as a coherent
injection that enhances rather than reverses the forward-pass
conversion.  The opposite phase choice produces coherent
back-conversion that partially undoes the forward-pass flattening of
$I(t)$.  That regime is not realized in commercial designs.  I
characterize the full $\delta\psi$ dependence, including this
back-conversion regime and the runaway band near $\delta\psi = \pi$,
in Sec.~\ref{sec:psisweep}.

\subsection{Nonlinear stage: split-step FFT computation}

The crystal stage $\hat{C}$ is computed in the time domain.  Given
the input mode vector $\{E_i^{(n)}\}$ at the crystal entrance, the
time-domain field $E(t)$ is constructed by inverse FFT.  The
coupled-wave equations~(\ref{eq:A3}) are then integrated through the
forward pass, the boundary condition $\delta\psi$ is applied at the
high-reflector, and the back-pass coupled-wave equations are
integrated.  The output IR field is projected back onto mode amplitudes
by forward FFT.  This split-step approach captures all pairwise
four-wave-mixing interactions among modes exactly within the mode
basis, with computational cost $O(N \log N)$ per round trip,
enabling simulation of $N \sim 200$ modes over multimillisecond
timescales.

The two measured observables are computed directly from the round-trip
green spectrum $\{E_{{\rm grn},m}^{(n)}\}$.  The full and
Fabry-Perot-filtered green powers on round trip $n$ are
\begin{equation}
\begin{split}
P_{\rm full}(n) &= \sum_m \bigl|E_{{\rm grn},m}^{(n)}\bigr|^2, \\
P_{\rm FP}(n)   &= \sum_m T_{\rm FP}(\omega_m)\,
                   \bigl|E_{{\rm grn},m}^{(n)}\bigr|^2,
\end{split}
\label{eq:observables}
\end{equation}
where $T_{\rm FP}(\omega_m)$ is the measured Fabry-Perot power
transmission at the green mode frequency $\omega_m$.  The RIN spectra
and RMS values reported in the text are computed from these
per-round-trip power series, with no further approximation.

\subsection{Noise sources}\label{app:noise}

Two noise mechanisms are included in the simulation.  Spontaneous
emission is modeled as additive complex Gaussian noise injected once
per mode per round trip:
\begin{equation}
\xi_i^{(n)} = \sqrt{R_{\rm sp}/2}\,
              \bigl(\xi_i^{(n,R)} + i\xi_i^{(n,I)}\bigr),
\label{eq:A4}
\end{equation}
where $\xi_i^{(n,R)}, \xi_i^{(n,I)} \sim \mathcal{N}(0,1)$ are
independent standard normal variates and
$R_{\rm sp} = n_{\rm sp} h\nu/\tau_c$ is the spontaneous emission
rate per mode, with $n_{\rm sp} = 2.0$ the inversion factor, $h\nu$
the photon energy, and $\tau_c$ the round-trip time.  Spontaneous
emission carries phase noise at the quantum-limit floor.

Technical noise is modeled as accumulating filtered fluctuations on
the saturated gain coefficient.  The gain coefficient is held
constant for $\Delta n$ round trips.  It is then updated by an additive
increment $\delta g$, scaled by $\sigma_{\rm gain}\sqrt{\Delta n}$.  The
increment is drawn from a Gaussian sequence with power spectral density
$\propto f^{-1.08}$, generated by FFT-shaping a white-noise sequence to
the target spectral slope.  The batched-update
interval $\Delta n$ reflects the timescale separation between the
round-trip period ($\tau_c \approx 2.8$~ns) and the much slower
technical-noise sources being modeled (pump fluctuations, thermal
drift, and mechanical noise, all of which vary on
microsecond-to-second timescales).  Successive batches accumulate, so
the gain trajectory is the running sum of the filtered increments
rather than a multiplicative perturbation; the resulting power
spectrum on the gain coefficient itself has a steeper low-frequency
rolloff than the per-batch increment spectrum.  The technical noise
drive amplitude is set by $\sigma_{\rm gain}$.

The technical noise has partial-common-mode mode-spatial structure
across the $N$ active modes.  Fifty independent noise streams are
generated, and mode $i$ draws from stream $i \bmod 50$.  Adjacent
modes therefore draw from different streams and are statistically
independent, while modes whose indices differ by a multiple of 50
share a stream and see identical gain noise.  This gives the technical
noise fifty independent degrees of freedom spread across the active
band, a partial-common-mode structure intermediate between fully
common-mode and fully independent injection.  It is a more realistic
representation of pump-speckle, thermal-lens, and mechanical noise
sources than either limiting case.  The technical-noise injection acts on the
gain coefficient (an amplitude-only perturbation).  A separate
technical-noise channel acting directly on individual mode phases
was tested across four decades of $\sigma_\varphi$ and is rejected
by the $\chi^{(2)}$ phase-organizing channel of
Sec.~\ref{sec:two-channels} in the resolvable spectral band of the
simulation.  Production runs therefore use $\sigma_\varphi = 0$ for
this technical-phase-noise channel as the operating choice.  This is
distinct from the quantum-limited phase noise carried by the
spontaneous-emission injection (Eq.~\ref{eq:A4}), which remains
active in all production runs and continues to drive the mode-phase
dynamics.  The $\sigma_\varphi = 0$ choice represents a
simulation-level finding about the technical-noise channel rather
than a determination that experimental phase-noise sources are
absent.

\subsection{Parameter values}

All simulations use parameters matched to the platform studied here.
The cavity length is taken to be 42~cm based on direct measurement of
the IR longitudinal beat-note frequency at 358~MHz (corresponding to
free spectral range $c/2L = 358$~MHz).

\begin{table}[h]
\centering
\caption{Simulation parameters used in all production runs.
$\sigma_{\rm gain}$ was calibrated against the full-output
RMS reported in Sec.~\ref{sec:vbprimary}; all other values are
fixed at their nominal hardware-derived numbers.  $N$ is the full mode
band above quantization noise; $N_{\rm active}$ counts modes with mean
power above $1\%$ of the peak.}
\label{tab:params}
{\small
\begin{tabular}{lll}
\hline
\hline
Quantity & Symbol & Value \\
\hline
Cavity length            & $L$              & 42 cm \\
Free spectral range      & $c/2L$           & 358 MHz \\
Round-trip time          & $\tau_c$         & 2.8 ns \\
Gain medium              & --                & Nd:YVO$_4$ \\
Fluorescence lifetime    & $\tau_f$         & 90 $\mu$s \\
Emission cross-section   & $\sigma_{\rm em}$ & $25 \times 10^{-19}$ cm$^2$ \\
Gain medium length       & $L_{\rm rod}$    & 5 mm \\
Fill factor              & $\eta$           & 0.012 \\
Small-signal gain        & $g_0$            & 4.0 \\
Round-trip linear loss   & $\ell$           & 0.25\% \\
Saturation power         & $I_s$            & 1.6 W \\
Rod position             & $z_0/L$          & 0.060 ($z_0 = 0.025$~m) \\
Nonlinear crystal        & --                & LBO, type I NCPM \\
Crystal length           & $L_c$            & 10 mm \\
Nonlinear coupling       & $\varepsilon$    & $3.84\times10^{-4}$ \\
High-reflector phase     & $\delta\psi$     & $2\pi$ \\
Full mode band           & $N$              & $\sim 200$ \\
Participating modes       & $N_{\rm active}$ & $\sim 86$ \\
Spontaneous emission $n_{\rm sp}$ & $n_{\rm sp}$ & 2.0 \\
Tech. noise drive        & $\sigma_{\rm gain}$ & $5\times10^{-6}$ \\
Tech. noise spectral slope & --              & $-1.08$ \\
Tech. noise update interval & $\Delta n$    & 50 round trips \\
Independent noise streams  & --              & 50 \\
\hline
\hline
\end{tabular}
}
\end{table}

The simulation is implemented as a C extension to a Python driver,
using FFTW3 for the time-domain FFT and integration steps.  The
standard simulation grid uses 1024 FFT points per round trip; the
green-filter diagnostic of Fig.~\ref{fig:nfsweep} uses a finer grid of
4096 points to resolve the narrow Fabry-Perot passband.
Typical runs use a warmup of $2.5 \times 10^5$ round trips (several
fluorescence lifetimes $\tau_f$) to reach the converged steady state.
The measurement record that follows is sized to the observable: of
order $10^5$ round trips for the microsecond time-domain figures, and
$5$~ms records (about $1.8 \times 10^6$ round trips) for the
low-frequency RIN spectra and the mode-count sweep of
Fig.~\ref{fig:nfsweep}.  Convergence with respect to FFT grid size and
warmup duration was verified by doubling each independently.

\section{Proof that $M_4$ is a Lyapunov functional for the forward
coupled-wave $\chi^{(2)}$ pass}\label{app:proof}

Section~\ref{sec:lyap} states that $M_4 \equiv
\langle I^2\rangle/\langle I\rangle^2$ decreases monotonically with
propagation distance through the crystal, with equality if and only if
the intracavity intensity $I(t)$ is constant.  The elementary
leading-order argument that establishes the descent in the
weak-conversion limit is given first, in Sec.~\ref{app:leading}; the
remainder of this appendix gives the all-orders single-pass result,
which uses the exact
$\mathrm{sech}^2/\tanh^2$ coupled-wave solution and the Chebyshev
integral inequality for co-monotone functions to extend the descent
beyond the weak-conversion limit.  The proof is rigorous for the
forward pass through the crystal under the zero-incident-second-harmonic
boundary condition and supplies a peak single-pass conversion bound
$\eta_{\rm peak} < 0.70$ (Sec.~\ref{app:co-mono}).  The application to
the back pass under the standard commercial dichroic geometry is
treated as an assumption supported by the simulation and experimental
results in the body, in which the descent operates per round trip
across the parameter regime studied.  The proof operates on the
time-domain intensity $I(t)$ and requires no assumption on mode count
or on individual mode amplitudes and phases.

\subsection{Leading-order proof}
\label{app:leading}

In the weak-conversion limit the descent follows from an elementary
argument.  The dimensionless coupling strength $\varepsilon \equiv
2\omega^2 d^2_{\rm eff} L_c^2 I_{\rm typ} / (n^3 \varepsilon_0 c^3)$
characterizes the $\chi^{(2)}$ interaction strength, with $d_{\rm eff}$
the effective nonlinear coefficient, $L_c$ the crystal length, $\omega$
the fundamental angular frequency, $I_{\rm typ}$ a typical intracavity
intensity, and the rest standard.
In this limit the $\chi^{(2)}$ depletion of the fundamental field
across the crystal is, to first order in $\varepsilon$,
\begin{equation}
E'(t) = E(t)\bigl[1 - \varepsilon |E(t)|^2 \bigr],
\label{eq:E-prime-leading}
\end{equation}
where the field on the left is the field after the crystal pass,
reconstructed from its Fourier components in the mode basis.  The
corresponding intensity is
\begin{equation}
I'(t) = I(t)\bigl[1 - 2\varepsilon I(t)\bigr] + O(\varepsilon^2).
\end{equation}
Each time slice loses energy in proportion to its own intensity
squared.  Bright slices are depleted more than dim slices, and the
structural form $-\varepsilon I(t)^2$ of the loss is what makes $M_4$
monotone.

To first order in $\varepsilon$, the moments $A = \langle I\rangle$ and
$B = \langle I^2\rangle$ update to
\begin{equation}
A' = A - 2\varepsilon B, \qquad
B' = B - 4\varepsilon \langle I^3\rangle,
\end{equation}
where primes denote values after the crystal pass.  Substituting into
$M_4' = B'/A'^2$ and expanding to first order in $\varepsilon$,
\begin{equation}
\Delta M_4 \equiv M_4' - M_4
   = -\frac{4\varepsilon}{A^3}\bigl[\langle I^3\rangle\langle I\rangle
                              - \langle I^2\rangle^2\bigr]
       + O(\varepsilon^2).
\label{eq:dm4-leading}
\end{equation}
The Cauchy-Schwarz inequality applied to $f = I^{3/2}$ and $g = I^{1/2}$
on the round-trip time average gives $\langle I^2\rangle^2
= \langle I^{3/2} I^{1/2}\rangle^2 \le \langle I^3\rangle\langle I\rangle$,
so the bracket in (\ref{eq:dm4-leading}) is nonnegative, and
$\Delta M_4 \le 0$, with equality if and only if $I(t)$ is constant.
The descent at leading order rests on three ingredients only: the
depletion $\delta I \propto -I^2$ from the $\chi^{(2)}$ interaction, the
Cauchy-Schwarz inequality on the round-trip time average, and no
assumption on mode count, individual mode amplitudes, or phases.  It
establishes the descent from the intensity-dependent depletion without
separating diagonal and off-diagonal contributions in the mode basis;
the amplitude-equalization and phase-organization channels of
Sec.~\ref{sec:two-channels} interpret that map physically.  The
remainder of this appendix removes the weak-conversion restriction.

\subsection{Setup}

Let the total intracavity field at the entrance of the crystal be
\begin{equation}
E(t) = \sum_j A_j \exp\!\bigl[i(\omega_j t + \varphi_j)\bigr],
\end{equation}
with mode amplitudes $A_j$ and phases $\varphi_j$ at frequencies
$\omega_j = \omega_0 + j\Omega$.  Define the instantaneous fundamental
intensity entering the crystal,
\begin{equation}
I_0(t) \equiv |E(t)|^2.
\end{equation}
The time average $\langle\cdot\rangle$ is taken over one round-trip
period $T = 2\pi/\Omega$.  The fourth-moment ratio is
\begin{equation}
M_4 = \langle I^2\rangle / \langle I\rangle^2.
\end{equation}
At each fixed $z$ I will write $A(z) = \langle I(t, z)\rangle$
and $B(z) = \langle I(t, z)^2\rangle$, with the average taken over
$t$ as above, so that $M_4(z) = B(z)/A(z)^2$.

\subsection{Exact coupled-wave solution}\label{app:exact}

Within the crystal, the fundamental field $E$ and second-harmonic
field $E_{\rm grn}$ evolve under the coupled-wave equations
(\ref{eq:A3}) of Appendix~\ref{app:model}.  For perfect phase matching and
on timescales short compared to the round-trip period (so that
each time slice $t$ of the input waveform $I_0(t)$ propagates through
the crystal independently), the equations admit the exact solution
\begin{align}
I(t,z)   &= I_0(t)\, \mathrm{sech}^2\bigl(\kappa z\sqrt{I_0(t)}\bigr),\nonumber\\
S(t,z)   &= I_0(t) - I(t,z)
          = I_0(t)\, \tanh^2\bigl(\kappa z\sqrt{I_0(t)}\bigr),
\label{eq:B1}
\end{align}
where $\kappa$ is the nonlinear coupling coefficient of
Appendix~\ref{app:model} and $S(t,z)$ is the second-harmonic power
generated up to position $z$.  Energy conservation $I + S = I_0$
holds identically.  The local SHG rate is
\begin{equation}
\begin{split}
S'(t,z) &\equiv \frac{\partial S}{\partial z}
   = -\frac{\partial I}{\partial z} \\
   &= 2\kappa\, I(t,z)\, \sqrt{I_0(t)}\,
       \tanh\!\bigl(\kappa z\sqrt{I_0(t)}\bigr) \ge 0.
\end{split}
\end{equation}

The solution~(\ref{eq:B1}) holds rigorously for the forward pass
under the boundary condition $E_{\rm grn}(z=0,t) = 0$ at the crystal
entrance.  This means no second-harmonic field is incident on the
crystal at the start of the propagation.  The forward-pass descent of $M_4$
established in Sec.~\ref{app:chebyshev} is therefore a rigorous
result of this appendix.

For the back pass through the crystal in the dichroic double-pass
geometry described in Appendix~\ref{app:model}, the boundary
condition is modified to $E_{\rm grn}(z=L_c,t) = E_{\rm grn}^F(L_c,t)
\exp(i\delta\psi)$, where $E_{\rm grn}^F$ is the green field at the
crystal exit at the end of the forward pass and $\delta\psi$ is the
relative phase shift between IR and green imposed by the
high-reflector.  With nonzero incident green, the coupled-wave
equations no longer admit the simple $\mathrm{sech}^2/\tanh^2$
solution~(\ref{eq:B1}); the back-pass dynamics depend on the phase
relationship between the incident green and the IR.  I assume that,
for the standard commercial constructive-interference choice of
$\delta\psi$ in which both passes deplete IR and add to green, the
back-pass map preserves the order-preserving structure that drives
the descent.  Under this assumption the descent applies separately
to each pass, giving $M_4^{\rm out} \le M_4^{\rm after\text{-}forward}
\le M_4^{\rm in}$.  The simulation results presented in body
Sec.~\ref{sec:psisweep} show that the round-trip descent holds across
a broad band of $\delta\psi$ around the operating point.  It fails only
near $\delta\psi = \pi$, where the back-reflected green re-enters in
antiphase and the back-pass $\chi^{(2)}$ becomes net parametric gain
rather than a dissipation.  The descent holding where the back pass is
dissipative, and failing where it is not, is consistent with the
order-preserving assumption above.  A formal extension of the analytic
proof to general $\delta\psi$ is left to future work.

\paragraph*{Per-time-slice independence.}
The exact time-domain solution~(\ref{eq:B1}) treats each temporal
slice $t$ of the input waveform as propagating independently through
the crystal.  This requires more than timescales short compared to
the round-trip period: it also requires that group-velocity
mismatch, group-velocity dispersion, finite phase-matching bandwidth,
and walkoff do not couple neighboring time slices over the relevant
bandwidth.  For the Nd:YVO$_4$/LBO system at the operating point of
this work, these conditions are well satisfied: the crystal length
$L_c = 10$~mm corresponds to a group-velocity-mismatch walkoff time
of $\sim 3$~ps between IR and green, much shorter than the inverse
intermode beat ($\sim 3$~ns) over which the multimode waveform
varies; the multimode IR bandwidth $N \cdot \Delta\nu \approx 200
\times 358$~MHz $\approx 72$~GHz $\approx 2.4$~cm$^{-1}$ is well
below the type~I noncritical phase-matching bandwidth of LBO
($\sim 50$~cm$^{-1}$); and spatial walkoff is negligible at noncritical
phase matching by construction.  The proof therefore applies to the
relevant operating regime without correction, although the
per-time-slice approximation should be reassessed if the framework
is extended to femtosecond pulses, broader IR bandwidths, or
critical phase-matching geometries.

\subsection{Computing $dM_4/dz$}

Differentiating $M_4 = B/A^2$ with respect to $z$ and using
$\partial I/\partial z = -S'$,
\begin{align}
\frac{dA}{dz} &= -\langle S'\rangle, \nonumber\\
\frac{dB}{dz} &= -2\langle I S'\rangle,
\end{align}
so that
\begin{align}
\frac{dM_4}{dz}
   &= \frac{1}{A^2}\frac{dB}{dz}
      - \frac{2B}{A^3}\frac{dA}{dz} \nonumber \\
   &= \frac{2}{A^3}\bigl[\langle I^2\rangle\langle S'\rangle
                       - \langle I\rangle\langle I S'\rangle\bigr].
\end{align}
Equivalently, using $B = \langle I^2\rangle$ and $A = \langle I\rangle$,
\begin{equation}
\frac{dM_4}{dz} = \frac{2}{A^3}\bigl[B\langle S'\rangle
                                    - A\langle I S'\rangle\bigr].
\label{eq:B2}
\end{equation}
The sign of $dM_4/dz$ is determined by the sign of $B\langle S'\rangle
- A\langle I S'\rangle$.

\subsection{Co-monotonicity in $I_0$}\label{app:co-mono}

At each fixed $z > 0$, both $I(t,z)$ and $S(t,z)$ are functions of
$I_0(t)$ alone, since each time slice evolves independently through
the crystal.  I need that $I$ is monotonically increasing in $I_0$
within the regime of validity stated below.  Co-monotonicity of
$S'/I$ with $I$ is established alongside the Chebyshev step in
Sec.~\ref{app:chebyshev}.

Writing $f(x) = x^2 \mathrm{sech}^2(\kappa z x)$ with $x = \sqrt{I_0}$
so that $I = f(x)$,
\begin{equation}
f'(x) = 2x\, \mathrm{sech}^2(\kappa z x)\,
          \bigl[1 - \kappa z x\, \tanh(\kappa z x)\bigr].
\end{equation}
This is positive when $\kappa z x \tanh(\kappa z x) < 1$.  The
transcendental equation $u\tanh u = 1$ has solution $u_c \approx
1.200$.  The condition therefore reads $\kappa z\sqrt{I_0} < u_c$
at every $(t, z)$ within the crystal.  This is a condition on the peak
instantaneous intensity in the waveform, $I_{0,{\rm max}} =
\max_t I_0(t)$.  The corresponding peak single-pass conversion
efficiency at the brightest time slice is
\begin{equation}
\eta_{\rm peak} = \tanh^2\!\bigl(\kappa L\sqrt{I_{0,{\rm max}}}\bigr)
                  < \tanh^2(u_c) \approx 0.695,
\end{equation}
equivalently approximately 70\%.  Below this bound, $I$ is
monotonically increasing in $I_0$ at every position $z$ in the
crystal.

The bound is on peak conversion at the brightest time slice, not on
time-averaged conversion.  The peak-to-average ratio of a multimode
waveform depends on its statistics: for a flat (constant-intensity)
waveform the two are equal, while for a spiky waveform far from the
manifold the peak can substantially exceed the average.  For
laboratory intracavity-doubled solid-state lasers the time-averaged
single-pass conversion efficiency is typically a few percent
($\approx 2$\% for the system studied here), so even peak-to-average
ratios of order 30 keep the peak conversion comfortably below the
0.70 bound.  Realistic multimode waveforms in the operating regime
do not approach the limit.

\subsection{Chebyshev integral inequality}\label{app:chebyshev}

For any two real-valued functions $f, g$ that are both monotonically
increasing (or both monotonically decreasing) in a common variable
and averaged with respect to a common probability measure $\mu$,
\begin{equation}
\int fg\, d\mu \ge \int f\, d\mu \cdot \int g\, d\mu,
\label{eq:B3}
\end{equation}
with equality if and only if $f$ or $g$ is constant $\mu$-a.e.  This
is Hardy-Littlewood-P\'olya, \textit{Inequalities} (Cambridge, 1952),
Theorem~236.

From~(\ref{eq:B2}), $dM_4/dz \le 0$ is equivalent to
\begin{equation}
A\langle I S'\rangle \ge B\langle S'\rangle.
\label{eq:B4}
\end{equation}
Define the intensity-weighted probability measure on one round-trip
period,
\begin{equation}
d\mu_I = I\, dt / (T A), \qquad \int d\mu_I = 1,
\end{equation}
under which
\begin{align}
\mathbb{E}_{\mu_I}[I]   &= \langle I^2\rangle/\langle I\rangle = B/A,\nonumber\\
\mathbb{E}_{\mu_I}[S'/I] &= \langle S'\rangle/A,\nonumber\\
\mathbb{E}_{\mu_I}[I\cdot S'/I] &= \langle I S'\rangle/A.
\end{align}
Consider the pair $(I, S'/I)$.  The first factor is increasing in
$I_0$ (Sec.~\ref{app:co-mono}).  The second factor is also increasing
in $I_0$: substituting the $\mathrm{sech}^2/\tanh^2$ solution and
writing $u = \kappa z\sqrt{I_0}$,
\begin{equation}
S'(t,z)/I(t,z) = 2\kappa\sqrt{I_0}\,\tanh(\kappa z\sqrt{I_0}),
\end{equation}
which is monotonically increasing in $I_0$ for all $z > 0$.  The
ratio $S'/I$ is undefined where $I = 0$; this set has measure zero
under $d\mu_I = I\, dt/A$, and the limiting expression
$2\kappa\sqrt{I_0}\,\tanh(\kappa z\sqrt{I_0})$ is continuous and
bounded on the relevant domain, so the inequality below holds
without further qualification.  Both factors are co-monotone in
$I_0$ at fixed $z$, so the Chebyshev inequality~(\ref{eq:B3})
applied to $(I, S'/I)$ under $d\mu_I$ gives
\begin{equation}
\mathbb{E}_{\mu_I}[I\cdot S'/I] \ge
   \mathbb{E}_{\mu_I}[I]\cdot\mathbb{E}_{\mu_I}[S'/I],
\end{equation}
i.e.
\begin{equation}
\langle I S'\rangle/A \ge (B/A)(\langle S'\rangle/A),
\end{equation}
which rearranges to~(\ref{eq:B4}).  Substituting into~(\ref{eq:B2}),
\begin{equation}
\frac{dM_4}{dz} \le 0 \quad \text{for every } z \in [0, L_{\rm crystal}],
\label{eq:B5}
\end{equation}
with strict inequality at every $z > 0$ unless $I(t,z)$ is constant
in $t$.  At $z = 0$ the rate $S'$ vanishes identically and the
inequality is satisfied trivially without any constraint on the
waveform; the descent statement therefore applies nontrivially to
the integrated crystal pass $z \in (0, L_{\rm crystal}]$ and to
$dM_4/dz$ at any interior $z > 0$.

\subsection{Scope and remarks}

\paragraph*{Scope.}
The result~(\ref{eq:B5}) holds for arbitrary mode amplitudes and
phases at the input to the crystal, at every position $z$ within the
forward pass, requiring only that the peak instantaneous single-pass
conversion efficiency satisfy
$\eta_{\rm peak} = \tanh^2(\kappa L\sqrt{I_{0,{\rm max}}}) < 0.70$.
No restriction on mode count, no assumption of weak coupling, and no
prior assumption on phase organization is needed.  The bound is on
peak rather than time-averaged conversion.  For laboratory
intracavity-doubled solid-state lasers operating at time-averaged
conversion of a few percent, even peak-to-average ratios of order 30
leave the peak conversion comfortably below the bound, so the proof
applies with substantial headroom.

\paragraph*{Role of back-conversion.}
The co-monotonicity of $I(t,z)$ and $S(t,z)$ in $I_0(t)$ is a direct
consequence of the $\mathrm{sech}^2$ structure of the exact
coupled-wave solution (Sec.~\ref{app:exact}).  The leading-order
proof of body Sec.~\ref{sec:lyap} also follows from this solution by
truncation: the field depletion $E'(t) = E(t)[1 - \varepsilon|E(t)|^2]$
[Eq.~(\ref{eq:E-prime-leading})] is the first-order Taylor expansion of the
$\mathrm{sech}^2/\tanh^2$ form in $\varepsilon$, equivalent to
discarding back-conversion.  Truncation breaks the co-monotonicity of
$I$ and $S$ beyond leading order, which is why the Cauchy-Schwarz
argument of body Sec.~\ref{sec:lyap} succeeds only in the
weak-conversion limit while the all-orders descent established here
requires the Chebyshev argument on the full $\mathrm{sech}^2$
solution.  The back-conversion channel of the $\chi^{(2)}$
interaction is therefore essential to the all-orders result, not as
a perturbative correction but as the structural feature that closes
the proof.

\paragraph*{Relation to Boltzmann's H-theorem.}
The result is structurally analogous to Boltzmann's
H-theorem in kinetic theory: a scalar functional of the dynamical
variables decreases monotonically under the dynamics until it reaches
its extremum on a manifold of equilibrium states.  Here $M_4$ plays
the role of $H$, the crystal plays the role of the collision
operator, and the $M_4 = 1$ manifold of constant-intensity multimode
configurations plays the role of the Maxwell-Boltzmann equilibrium.
The Chebyshev inequality plays the role that Jensen's inequality
plays in the original H-theorem.  The analogy is formal only: it is a
monotone-descent structure, not a thermodynamic one, and $M_4$ is not
an entropy.

\paragraph*{Round-trip behavior.}
Equation~(\ref{eq:B5}) is a per-pass, in-crystal result: $M_4$ is
nonincreasing across each crystal traversal.  Global attraction of
the $M_4 = 1$ manifold under the full round-trip dynamics is not
proved by the Chebyshev result alone.  Those dynamics include gain,
SHB cross-saturation, noise sources, and inter-pass mode coupling.  It
is supported by the simulation evidence presented in the body, in
which the round-trip composition empirically reaches the manifold
from arbitrary initial conditions across the parameter regime
studied.  Between crystal passes the gain medium and noise sources
can drive $M_4$ back up.  The observed steady-state $M_4$ is the
round-trip balance between the in-crystal descent and the
inter-pass drive (body Sec.~\ref{sec:roundtrip}).

\paragraph*{Agnosticism to input phase configuration.}
The proof depends on the structure of the $\chi^{(2)}$ amplitude
depletion and on monotonicity of $I$ in $I_0$ within the validity
range, but it makes no assumption about how the input waveform
$I_0(t)$ was produced.  Whether the input mode amplitudes and phases
were generated by $\chi^{(2)}$ phase coupling on previous round
trips, by other phase-affecting mechanisms in the cavity
(gain-medium dynamics, dispersion, thermal effects), or by random
fluctuations, the per-pass crystal traversal reduces $M_4$ in the
operating regime.  The steady-state $M_4 \approx 1$ observed
experimentally is therefore the round-trip balance between the
in-crystal descent and all other dynamics combined, including any
phase-affecting mechanisms beyond those explicitly modeled in this
work.  This robustness is a structural feature of the proof, not a
numerical observation.

\section{History, prior treatments, and related lasers}
\label{app:prior}

The intracavity-doubled multimode laser was analyzed by me
through rate equations~\cite{baer1986}, identifying the
$\chi^{(2)}$ mode-coupling mechanism that produces bistability,
mode-hopping, and chaotic dynamics at $N = 2$ and $N = 3$.  The
framework was developed by others over four decades
\cite{wumandel1987,wiesenfeld1990,roybracikowskijames1991,erneuxmandel1995,kozyreffmandel1998,pietrzykdanailov2000},
remaining within rate-equation variables tracking mode intensities
and population inversions.  The original analysis was developed for
and validated against few-mode operation.  The present treatment
includes complex mode amplitudes including phase, identifies the
Lyapunov descent of body Sec.~\ref{sec:lyap}, and addresses the
high-mode-count, low-noise regime where most commercial
intracavity-doubled lasers operate.

\begin{figure*}[tb]
\centering
\includegraphics[width=\textwidth]{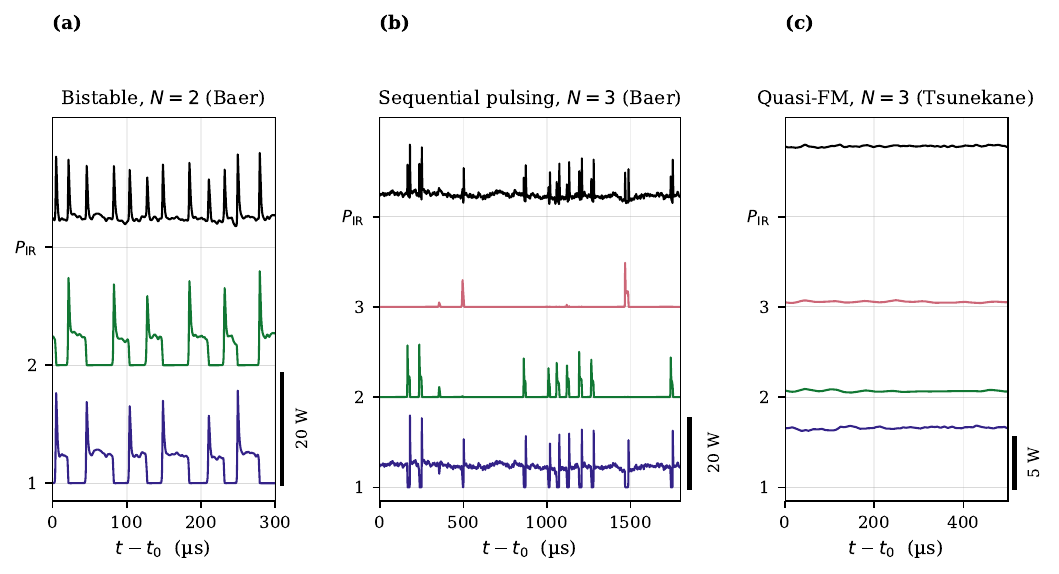}
\caption{Few-mode and historical regimes reproduced by the model, in an
offset/waterfall layout: per-mode circulating IR power $|E_j(t)|^2$ each
in its own lane (numbered by mode index $j$) with the sum $P_{\rm IR}$
stacked on top; the per-panel bar gives the power scale.  (a)~$N=2$ bistable mode-hopping (Baer 6~cm
cavity, etalon, $z_0 = L/30$, seed~42): the two modes hop anti-phase.
(b)~$N=3$ sequential pulsing (Baer 6~cm cavity, etalon, seed~17): the
modes pulse aperiodically in turn.  (c)~$N=3$ quasi-FM (Tsunekane 60~cm
cavity, birefringent filter, $z_0 = L/3$, seed~42): a carrier plus two
sidebands, a stationary state near the manifold.  Panels~(a)--(c) use
minimal seed noise ($\sigma_{\rm gain} = 1.7\times10^{-7}$) to display
the deterministic attractors cleanly.}
\label{fig:regimes}
\end{figure*}

The model reproduces the historical few-mode regimes of
intracavity-doubled lasers, shown in Fig.~\ref{fig:regimes} as the
per-mode circulating power $|E_j(t)|^2$ (each offset in its own lane)
with the sum $P_{\rm IR}$ stacked on top.  Panel~(a) is bistable
mode-hopping at $N = 2$ with the cavity etalon tuned for two-mode
operation, reproducing the bistable regime first reported by
Baer~\cite{baer1986}: the two modes alternate anti-phase in
spike-and-decay events and the total power swings periodically by
several times its mean (94\% green RMS).  Panel~(b) is sequential
pulsing at $N = 3$, reproducing the chaotic regime of
Ref.~\cite{baer1986}: the modes spike aperiodically with irregular
timing (35.0\% green RMS).  Panel~(c) is quasi-FM-quadrature locking at
$N = 3$ in the Tsunekane 1997 cavity ($z_0 = L/3$)
\cite{tsunekane1997,anthon1999}: a carrier near 6~W with two
$\sim 0.6$~W sidebands, a quasi-FM state near the constant-intensity
manifold (2.24\% green RMS).  The same Lyapunov structure, standing-wave
spatial-hole-burning kernel, and coupled-wave crystal physics produce
all of these together with the quiet high-mode-count regime of body
Fig.~\ref{fig:quiet}.  Panels~(a),(b) are off-manifold limit cycles
(the gain dynamics overpower the crystal descent); panel~(c), like the
quiet regime, is a near-manifold stationary state.

The structural distinction between the two frameworks is the
following.  Rate-equation models track $I_j = |E_j|^2$.  The
constant-intensity condition that defines the $M_4 = 1$ manifold is
$\sum_{jk} E_j E_k^* \exp[i(\omega_j - \omega_k)t] = {\rm const}$,
a statement about phase-coherent sums of complex mode amplitudes.
The off-diagonal terms are not present in intensity-only variables.
The Lyapunov descent of body Sec.~\ref{sec:lyap} operates on the time
waveform $I(t)$ that contains these off-diagonal beats.  The
mode-basis decomposition into amplitude-equalization and
phase-organization channels described in body
Sec.~\ref{sec:two-channels} identifies two structurally distinct
couplings acting on complex mode amplitudes.  The amplitude-
equalization channel is present in intensity-only treatments as the
SFG cross-loss term, and is sufficient on its own to drive the
leading-order Lyapunov descent (body Sec.~\ref{sec:lyap}).
Rate-equation treatments cannot access the phase-organization channel,
the coherent four-wave-mixing torques that operate on mode phases.  Nor
can they access the all-orders descent structure, which requires phase
information.  The Lyapunov framework as developed here
therefore overlaps with rate-equation dynamics at leading order in
the descent mechanism but extends beyond them in the all-orders
treatment and in the identification of the phase-organization
channel.

I have implemented the rate-equation system
\cite{baer1986,roybracikowskijames1991,pietrzykdanailov2000} with
adaptive integration and validated it against Baer's published
$N = 3$ chaotic dynamics (RMS 46\%, $M_4 = 1.21$).  The field model
also reproduces these few-mode dynamics with realistic
mode-dependent losses.  Applied at the typical operating parameters
of the system studied here with calculated SHB cross-saturation, the
rate-equation framework predicts severe mode-count reduction across
the physically plausible range of $\varepsilon$.  Intensity-only
treatments collapse to a small number of surviving modes rather
than sustaining the high-mode-count regime.  The reduction persists
when technical noise calibrated to the field-code parameters is
added.  At the same operating parameters with the same noise drive,
the field model sustains 200 active modes with intracavity (IR)
intensity RMS 0.18\% and $M_4 = 1.0003$.  The measured observable is
the green output, whose simulated RMS ($0.42 \pm 0.03\%$;
Sec.~\ref{sec:vbprimary}) agrees with the experimental
$0.44 \pm 0.02\%$ green RMS at $N \approx 200$ active
modes (Fig.~\ref{fig:fp} of the body).

This places the present treatment as the field-theoretic completion
of the program initiated in Ref.~\cite{baer1986}.  The $\chi^{(2)}$
mode coupling that produces few-mode chaos in rate-equation
variables is the same coupling that, when treated with complex mode
amplitudes including phase, organizes high-mode-count operation onto
the constant-intensity manifold.

\subsection{Contrast with single-mode nonlinear quieting}
\label{app:singlemode}

The mechanism is qualitatively distinct from passive nonlinear
intensity-noise suppression in single-mode systems.  In
single-frequency second-harmonic generation the nonlinearity acts as
an intensity-dependent loss that directly clamps the fluctuations of
the single field, with a noise reduction that grows with the
conversion efficiency~\cite{shg_noise_porat2024,shg_power_zheng2024}.
Related single-frequency schemes use nonlinear or quantum resources to
filter phase noise or generate squeezed
light~\cite{laser_phasenoise_li2026,squeezed_khz_li2025}.
Here the nonlinearity instead \emph{increases} each field's
fluctuations and achieves quieting only collectively, through the phase
organization of the mode ensemble, an effect with no single-mode
analogue.  Consistent with this collective origin, the total-output
suppression is robust to the coupling strength: over two decades of
$\varepsilon$ the per-mode noise and the cancellation tighten in step,
leaving the aggregate noise nearly unchanged, whereas single-mode
suppression improves monotonically with the coupling.

\subsection{Relation to the FM laser}\label{app:fm}

The closest prior amplitude-stable multimode laser is the FM laser
of Harris and Targ~1964 \cite{harristarg1964}.  There an intracavity
phase modulator is driven at the cavity round-trip angular frequency
$\Omega = \pi c / L_{\rm cav}$.  It forces the field into a
Bessel-amplitude state with phases locked such that $|E(t)|^2$ is
constant.  The constant-intensity condition can be expressed without
reference to the modulator: any periodic phase function $\varphi(t)$
for which $E(t) = A_0 \exp(i\varphi(t))$ has fixed amplitude defines
a point on the $M_4 = 1$ manifold.  Explicitly, the constant-intensity
condition $|E(t)|^2 = A_0^2$ requires $\varphi(t)$ to be a periodic
real function, parametrized by the Fourier expansion
\begin{equation}
\varphi(t) = \sum_{n=1}^\infty \beta_n \cos(n\Omega t + \alpha_n).
\end{equation}
The Bessel-amplitude FM state corresponds to $\beta_1 = \beta$ with
$\beta_n = 0$ for $n \ge 2$.  Other choices of $(\beta_n, \alpha_n)$
coefficients produce different mode-amplitude distributions but the
same constant-intensity property.  An exactly constant-modulus field
with more than one mode requires an unbounded sideband series, so at
finite mode count the constant-intensity condition is only approached,
with a residual $M_4 = 1 + \delta_N$ that shrinks as the available mode
band widens.  I therefore use ``the $M_4 = 1$ manifold'' as shorthand
for this near-constant-intensity limit, restricted to mode-amplitude
distributions consistent with the gain bandwidth.  The FM-Bessel state
is one point on it.

The FM laser and the Lyapunov descent reach this manifold through
different mechanisms.  The FM laser uses external coherent forcing.
The system response is a driven equilibrium with no requirement for
any internal stability mechanism.  In the He-Ne laser of Harris and
Targ the gain medium is a gas, so atomic motion and Doppler broadening
wash out the standing-wave intensity grating and spatial hole burning
does not resist the Bessel state.  The
$\chi^{(2)}$ Lyapunov descent of the body drives the system onto the
manifold from arbitrary initial conditions with no external driving.
The state reached is selected jointly by the descent and the
gain-SHB constraint surface.  Under standing-wave conditions with
strong SHB, this selection lies far from the Bessel point.  The reason
is SHB cross-saturation.  It depletes the gain more strongly at modes
with higher intracavity intensity, so it selects an amplitude envelope
that tracks the gain profile rather than the externally imposed Bessel
envelope of the FM laser~\cite{adamsmakerferguson1990}.  This holds for
both Nd:YAG and the Nd:YVO$_4$ laser studied here, at the operating
geometry of this work.

Gain-bandwidth truncation acts differently in the two systems.  In
an idealized FM laser the Bessel distribution extends infinitely in
mode index.  In any real laser the gain bandwidth limits the number
of modes carrying nonzero amplitude, producing a residual amplitude
modulation at integer multiples of $\Omega$ because the
Bessel-amplitude condition can no longer be satisfied exactly.  In
the Lyapunov descent, the same bandwidth limit restricts which
$\varphi(t)$ functions on the manifold are realizable.  The descent
and the gain-SHB constraint surface jointly select a $\varphi(t)$
within the bandwidth-allowed subset, and this selection is not
restricted to the Bessel or FM condition.  The mode-amplitude
distributions observed both in the simulation and in
spectrally resolved output of the system studied here generally do
not resemble an FM-Bessel envelope and are not stationary.  The
system explores points on the manifold rather than locking to a
single configuration.  Numerically the resulting amplitude residual
is at the $10^{-5}$ level for typical operating parameters at
$N \approx 200$.

The Anthon~1999 analysis~\cite{anthon1999} sits between these two
limits and merits explicit comment because of its proximity to the
picture developed here.  Anthon shows that the ideal FM state is compatible with minimizing
the nonlinear loss.  With no amplitude modulation the second-harmonic
output coupling is minimized, and placing the gain medium at
$z_0 = L/2$ with modulation depth $u = u_0/[2\sin(\pi z_0/L)]$
($u_0 \approx 2.405$ the first zero of $J_0$, so that
$J_0[2u\sin(\pi z_0/L)] = 0$) minimizes the spatial hole burning,
within a tolerance $\Delta z \le L[8\varepsilon/(\pi^2 u_0
J_1(u_0))]^{1/2}$.  This holds for the few-mode operating point of
Tsunekane, Taguchi, and Inaba~\cite{tsunekane1997}.  A real laser with
a finite, structured gain profile cannot meet the exact steady-state
FM condition, and with only a few modes the Bessel amplitudes are
unreachable, so a residual amplitude modulation remains.  Their
three-mode green laser is low-noise but not noiseless, with a residual
RIN at the $-130$~dB/Hz level, and the three-mode model here shows the
same feature: the Tsunekane-geometry state of Fig.~\ref{fig:regimes}(c)
is a quasi-FM state near the constant-intensity manifold, with $2.24\%$
residual green RMS, not the exact Bessel point.  Their detector
bandwidth ($\sim 20$~kHz, narrow compared to the intermode beat of
several hundred MHz) records this low-frequency residual but does not
resolve the fast $2\Omega$ modulation that would separate a truncated
quasi-FM state from a true Bessel state.  What Anthon does
not establish is that arbitrary initial conditions evolve onto the
FM-Bessel point.  Anthon adopts the FM ansatz for the steady state
and verifies its compatibility with the geometry, rather than
deriving the basin of attraction from the underlying coupled-mode
dynamics.  The Lyapunov framework supplies the missing piece.  The
crystal step has a Lyapunov descent toward the constant-intensity
manifold under the assumptions of Appendix~\ref{app:proof}; the full
round-trip laser approaches the manifold in the simulations and in
the measured operating regime.  The gain-SHB constraint surface
determines which point on the manifold the round-trip dynamics
actually select.  At symmetry-protected gain placements such as
$z_0 = L/3$ or $z_0 = L/2$ with appropriate modulation depth, the
constraint surface can intersect the manifold at or near the Bessel
point, recovering Anthon's identification as a special case.  At
generic placements with strong SHB the intersection lies in the
gain-shaped-envelope region of the manifold instead.

\section{Principal-component decomposition of the quiet state}
\label{app:pca}

This appendix gives the covariance decomposition summarized in Sec.~\ref{sec:pca} and the window-dependence of the loud subspace.

\begin{figure}[tb]
\centering
\includegraphics[width=\columnwidth]{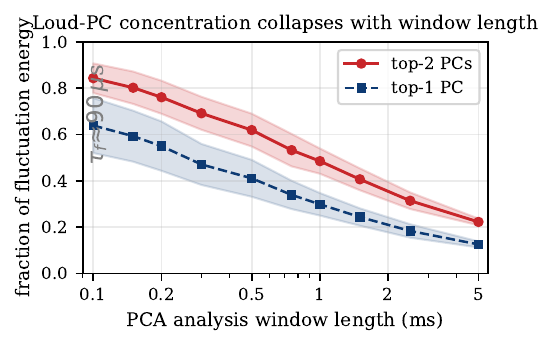}
\caption{How concentrated the noise looks in a few principal components
is an artifact of the PCA window length, not a fixed property.  Fraction
of the modal fluctuation energy carried by the largest one and two
principal components versus window length, averaged over eight noise
seeds and all nonoverlapping windows of each length (shaded band:
standard deviation).  Over a short window ($\ll \tau_f$) the loud
directions are quasi-frozen and the noise appears concentrated in one or
two principal components; over a window long compared with the
gain-recovery time $\tau_f \approx 90~\mu$s (dotted line) the loud
directions have reorganized many times and the noise spreads over many.
The principal components are collective combinations of the $\sim 200$
longitudinal modes, not individual modes; the quiet total-power
direction is unaffected.}
\label{fig:pcawindow}
\end{figure}

Let $\delta I_i(t) = I_i(t) - \langle I_i\rangle$ be the per-mode
intensity fluctuation about the mean.  The total fluctuation is
$\delta P_{\rm tot}(t) = \sum_i \delta I_i(t) = \sqrt{N}\,
\mathbf{u} \cdot \bm{\delta I}(t)$, where
$\mathbf{u} = (1,1,\dots,1)/\sqrt{N}$ is the unit vector along the
uniform direction in mode-amplitude space.  I compute the covariance
matrix $C$ of the per-mode fluctuations from a $10^4$-pass record
window of approximately 28~$\mu$s at the converged steady state.
This window is short compared with the mode-reorganization time
$\tau_f$, so it resolves the loud fluctuation directions as a
quasi-stationary snapshot before they drift; the dependence of the
decomposition on window length is examined below
(Fig.~\ref{fig:pcawindow}).
Diagonalizing $C$ gives eigenvalues $\lambda_n$ and eigenvectors
$\mathbf{v}_n$, the principal components of the mode-fluctuation
distribution.  The principal components form an orthogonal basis on
which the per-mode fluctuation pattern can be expanded, and
$\mathrm{var}(P_{\rm tot})$ decomposes into a sum of contributions
from the individual principal components.

Two scalar quantities determine each principal component's
contribution to $\mathrm{var}(P_{\rm tot})$.  The first is its
eigenvalue $\lambda_n$, which is the variance of the principal
component's time-domain projection $\xi_n(t) = \mathbf{v}_n \cdot
\bm{\delta I}(t)$.  This is the intrinsic fluctuation amplitude of
that principal component.  The second is its squared overlap with
$\mathbf{u}$, defined as $c_n^2 = (\mathbf{v}_n \cdot \mathbf{u})^2$,
which measures how strongly the principal component overlaps with the
experimental observable.  Because the principal components are
orthogonal and the projections $\xi_n(t)$ onto distinct eigenvectors
are uncorrelated, cross-terms in the variance of $P_{\rm tot}$
vanish, and the variance of $P_{\rm tot}$ decomposes as
\begin{equation}
\mathrm{var}(P_{\rm tot}) = N \sum_n c_n^2\, \lambda_n.
\label{eq:var-decomp}
\end{equation}
A principal component contributes little to
$\mathrm{var}(P_{\rm tot})$ if either factor is small.  Either it has
small intrinsic fluctuation, or its direction in mode space is nearly
orthogonal to $\mathbf{u}$ so its fluctuations cannot reach the
experimental observable.

Equivalently, a loud direction orthogonal to $\mathbf{u}$ is a
sum-zero pattern in the mode basis: the large mode fluctuations it
carries are anticorrelated and cancel in the total.  This is the same
cancellation that appears directly in
$\mathrm{var}(P_{\rm tot}) = N\,\mathbf{u}^{\top} C\,\mathbf{u}
= \sum_i C_{ii} + \sum_{i\neq j} C_{ij}$, where the large individual
variances $C_{ii}$ are nearly offset by negative off-diagonal
covariances.  Diagonalizing $C$ trades this mode-basis anticorrelation
for the orthogonality of the loud directions to $\mathbf{u}$: the
principal components do not cancel one another---being uncorrelated,
their variances add---but each reaches the observable only through its
small overlap $c_n$.

The loud subspace, by contrast, is neither low-dimensional nor
stationary, and its apparent structure depends on the length of the
analysis window.  Over a short window, of order the $28~\mu$s used for
the eigenvector basis of Fig.~\ref{fig:pca}, a single principal
component appears to dominate the fluctuation energy.  Over longer
windows the energy spreads over more components: the fraction carried by
the two largest principal components falls from about $0.7$ over
$0.3$~ms to about $0.2$ over $5$~ms (Fig.~\ref{fig:pcawindow}).  The
reason is that the loud collective modes drift.  The leading eigenvector
computed over one $0.2$~ms window is nearly orthogonal to the one
computed a millisecond later, and the leading eigenvalue itself varies
by tens of percent from window to window.  The characteristic time for
this reorganization is approximately $90$--$100~\mu$s.  I measure it
both as the half-decorrelation time of the leading noisy subspace and
as the autocorrelation time of the leading eigenvalue.

This time is the gain-recovery time $\tau_f$, and it identifies the
mechanism.  The
$\chi^{(2)}$ descent enforces the constant-intensity constraint
quickly, on the sub-microsecond per-pass timescale of the crystal, so
the total power is pinned flat almost immediately.  Motion \emph{within}
the constant-intensity manifold, however, is driven by the much slower
gain and spatial-hole-burning dynamics, which redistribute power among
the modes on the inversion-recovery time $\tau_f$.  Removing the
spontaneous-emission noise from the simulation leaves this drift time,
and the phase decorrelation of the modes, unchanged, confirming that the
reorganization is driven by the gain dynamics rather than by
spontaneous-emission phase diffusion.  The system therefore
diffuses slowly across a high-dimensional flat manifold while the total
output stays clamped: the mode combinations that momentarily carry the
noise are continually reorganized on $\sim\tau_f$, but their sum remains
quiet at all times.  A short analysis window catches the state
quasi-frozen along a few soft directions, so the noise appears
concentrated in one or two principal components; over a window long
compared with $\tau_f$ the loud directions reorganize many times and the
noise appears spread over many principal components.

\section{Detector and measurement characterization}\label{app:detector}

Both detection channels use identical detectors: the same photodiode,
transimpedance amplifier, and board layout.  The full-versus-filtered
contrast of Fig.~\ref{fig:fp} is independent of the detector response.
The detector bandwidth, measured directly by modulating a green LED,
exceeds 1~MHz, confirming that the sub-MHz RIN spectra of
Fig.~\ref{fig:fp}(e),(f) are not shaped by the detection electronics.  The
detector noise floor, measured with the beam blocked, is white
($\approx 1.8\times10^{-14}$~V$^2$/Hz) from a few~kHz to $\sim 100$~kHz
and rises above $\sim 200$~kHz through operational-amplifier noise-gain
peaking; a sensitive-scale baseline isolates the analog floor at
$\approx 2.8\times10^{-15}$~V$^2$/Hz.  The two channels share this
absolute floor, but the filtered channel transmits far less power.  Its
RIN floor therefore lies above the full-channel floor purely through
the $1/\langle V\rangle^2$ normalization of the relative intensity
noise, not through any physical difference between the channels.  At the operating point both channels are resolved well
above their floors across the frequency range of interest, so the
measured $\sim 100\times$ full-versus-filtered contrast reflects laser
dynamics rather than instrument noise.  At the coarse vertical scale
required to accommodate the DC level, the 8-bit oscilloscope
quantization contributes to the high-frequency floor of the quiet full
channel; the full-output noise reported here is therefore an upper
bound, and the true intensity-noise suppression may exceed it.

\bibliography{qmad_v25}

\end{document}